\documentclass{article}

\usepackage{microtype}
\usepackage{multirow}
\usepackage{soul}
\usepackage{mathtools}
\usepackage{amsmath,amsfonts, amsthm}
\usepackage{algorithmic}
\usepackage[ruled,vlined]{algorithm2e}
\usepackage{graphicx}
\usepackage{float}
\usepackage{lipsum}
\usepackage{subfig}
\usepackage{textcomp}
\usepackage{xcolor}
\usepackage{placeins}
\usepackage{hhline}
\usepackage{tabularx}
\usepackage{enumitem}
\usepackage{setspace}
\usepackage{booktabs}
\usepackage{adjustbox}
\usepackage[normalem]{ulem}
\usepackage{indentfirst}
\usepackage{titlesec}

\usepackage[accepted]{icml2023}

\usepackage{amsmath}
\usepackage{amssymb}
\usepackage{mathtools}
\usepackage{amsthm}

\usepackage[capitalize,noabbrev]{cleveref}

\usepackage[accepted]{icml2023}

\icmltitlerunning{Deep Graph Representation Learning and Optimization for Influence Maximization}

\theoremstyle{definition}

\newtheorem{dfn}{\textbf{Definition}}
\newtheorem{thm}{Theorem}

\newtheorem{cor}[thm]{Corollary}

\DeclareMathOperator*{\argmax}{arg\,max}

\let\norm\undefined 
\DeclarePairedDelimiter\norm{\lVert}{\rVert}

\icmltitlerunning{Deep Graph Representation Learning and Optimization for Influence Maximization}

\begin{document}

\twocolumn[
\icmltitle{Deep Graph Representation Learning and Optimization for Influence Maximization}



\icmlsetsymbol{equal}{*}

\begin{icmlauthorlist}
\icmlauthor{Chen Ling}{equal,emory}
\icmlauthor{Junji Jiang}{equal,fudan}
\icmlauthor{Junxiang Wang}{nec}
\icmlauthor{My Thai}{florida}
\icmlauthor{Lukas Xue}{emory}
\icmlauthor{James Song}{emory}
\icmlauthor{Meikang Qiu}{south}
\icmlauthor{Liang Zhao}{emory}
\end{icmlauthorlist}

\icmlaffiliation{emory}{Emory University, Atlanta, GA}
\icmlaffiliation{fudan}{Fudan University, Shanghai, CHINA}
\icmlaffiliation{nec}{NEC Labs America, Princeton, NJ}
\icmlaffiliation{florida}{University of Florida, Gainesville, FL}
\icmlaffiliation{south}{Dakota State University, Madison, SD}

\icmlcorrespondingauthor{Chen Ling}{chen.ling@emory.edu}
\icmlcorrespondingauthor{Liang Zhao}{liang.zhao@emory.edu}
\icmlkeywords{Machine Learning, ICML}

\vskip 0.3in
]



\printAffiliationsAndNotice{\icmlEqualContribution} 

\begin{abstract}
    Influence maximization (IM) is formulated as selecting a set of initial users from a social network to maximize the expected number of influenced users. Researchers have made great progress in designing various \textit{traditional} methods, and their theoretical design and performance gain are close to a limit. In the past few years, learning-based IM methods have emerged to achieve stronger generalization ability to unknown graphs than traditional ones. However, the development of learning-based IM methods is still limited by fundamental obstacles, including 1) the difficulty of effectively solving the objective function; 2) the difficulty of characterizing the diversified underlying diffusion patterns; and 3) the difficulty of adapting the solution under various node-centrality-constrained IM variants. To cope with the above challenges, we design a novel framework DeepIM to generatively characterize the latent representation of seed sets, and we propose to learn the diversified information diffusion pattern in a data-driven and end-to-end manner. Finally, we design a novel objective function to infer optimal seed sets under flexible node-centrality-based budget constraints. Extensive analyses are conducted over both synthetic and real-world datasets to demonstrate the overall performance of DeepIM. The code and data are available at: \url{https://github.com/triplej0079/DeepIM}.
\end{abstract}

\section{Introduction}
    As one of the fundamental research problems in network analysis, the objective of Influence Maximization (IM) is to find a set of seed nodes that maximizes the spread of influence in a social network. IM has been extensively studied in recent years due to its large commercial value. For example, consider the case of viral marketing~\cite{chen2010scalable} for promoting a commercial product, where a company may wish to spread the adoption of a new product from some initially selected users, the selected initial users are expected to spread the information about the product on their respective social networks. This cascading process will be continued, and ultimately, a significant fraction of the users will try the product. Besides viral marketing, IM is also the cornerstone in many other critical applications such as network monitoring~\cite{wang2017real}, misinformation containment~\cite{yang2020containment}, and friend recommendation~\cite{ye2012exploring}.

    As a typical type of combinatorial optimization problem, retrieving a (near) optimal seed set to maximize the influence in a network is challenging due to the stochastic nature of information diffusion and the hardness of the problem. Traditional (non-learning-based) methods for IM \cite{leskovec2007cost,kempe2003maximizing, tang2014influence,tang2015influence,nguyen2016stop,saito2012efficient} have made great progress in the last decade, and \citet{li2019tiptop} have even achieved exact solutions under specific diffusion models. The commonality of traditional methods is the explicit requirement of the information diffusion model as the model input. However, the real-world information diffusion process is complex and cannot be simply modeled by prescribed diffusion models. With the recent development of machine/deep learning, it is natural to consider a learning-based way to characterize the underlying diffusion process.

    While great progress has been made in the field, current efforts on learning-based IM solutions are still in the infancy stage due to fundamental obstacles as follows. \emph{1). The difficulty of efficiently optimizing the objective function.} Learning-based IM methods tend to solve the discrete problem in continuous space by mostly leveraging deep network representations \cite{zhang2022network, kumar2022influence} and deep reinforcement learning \cite{tian2020deep, li2022piano}. Even though they could attain a competitive performance with traditional methods, their scalability and execution efficiency are problematic due to \textit{(a)} the need to iteratively update all node embeddings at each action and \textit{(b)} the \#P-hardness of computing the influence spread \cite{lin2017boosting}. \emph{2). The difficulty of automatically identifying and modeling the actual diffusion process.} To maximize the influence spread in a network, the underlying information diffusion pattern is an imperative part as it determines the overall information propagation outcome. However, both traditional and learning-based methods cannot characterize the underlying diffusion process without heuristics. To work around this, both traditional and current learning-based methods have been leveraging pre-defined diffusion models (e.g., Linear Threshold (LT) and Independent Cascade (IC)) as the input to solve the combinatorial optimization problem. Although they could work well only for the process following their heuristics, the real-world network process is way more complex than the heuristics and largely unknown. \emph{3). the difficulty of adapting solutions to various node-centrality-constrained IM problems.} There are a lot of variants of IM that relate to node centrality, e.g., the constraint on the number of seed nodes, the constraint on the total degree of seed nodes, etc. Current learning-based IM solutions do not have a well-defined paradigm for solving different node-centrality-constrained IM problems, which poses another challenge to their solution adaptivity.
         	
    To address the above challenges, we propose a novel framework - DeepIM, to solve the IM problem by developing a novel strategy that embeds the initial discrete optimization domain into a larger continuous space. Remarkably, we propose to learn the latent representation of seed sets by retaining their expressiveness and directly optimizing in the continuous space to reduce the problem's hardness. We further design a learning-based diffusion model to characterize the underlying diffusion dynamics in an end-to-end manner. Moreover, we develop a generic seed set inference framework to directly optimize and generate set embeddings under a uniform budget constraint. Finally, we summarize our contributions as follows:
    \vspace{-3mm}
    \begin{itemize}[leftmargin=*]
      \setlength\itemsep{0em}
        \item \textbf{Problem.} We formulate the learning-based IM problem as embedding the initial discrete optimization domain into continuous space for easing the optimization and identify its unique challenges arising from real applications.
        \item \textbf{Framework.} We propose modeling the representation of the seed set in a latent space, and the representation is jointly trained with the model that learns the underlying graph diffusion process in an end-to-end manner.  
        \item \textbf{Adaptivity.} We propose a novel constrained optimization objective function to infer the optimal seed set by leveraging deep graph embeddings, which can be applied under arbitrary node-centrality-related constraints.
        \item \textbf{Evaluation.} We conduct extensive experiments over four real-world datasets to demonstrate the performance of the proposed method. Compared with other state-of-the-art in various application scenarios, DeepIM achieves the best results in finding a seed set to maximize the influence. 
    \end{itemize}
    
\section{Related Work}\label{sec:related}
    \subsection{Learning-based Influence Maximization}
    Influence Maximization (IM) was first formulated as a combinatorial optimization problem by \citet{kempe2003maximizing}, which has inspired extensive research and applications in the next decade. Most of the traditional (i.e., non-learning-based) IM methods can be categorized as simulation-based, proxy-based, and heuristic-based. Traditional methods have achieved near or exact solutions under specific diffusion models with efficiency. Note that \citet{du2014influence,vaswani2017model} have alluded to the possibility of learning the influence from the cascade data; however, they still assumed a prescribed model guides the diffusion pattern, specifically the Coverage function. We refer readers to recent surveys \cite{li2018influence,banerjee2020survey} for more detailed reviews of traditional methods.

    Learning-based methods use deep learning to address the drawbacks of traditional IM methods, namely the lack of generalization ability. Pioneered works \cite{lin2015learning, ali2018boosting} first combined reinforcement learning with IM, and they triggered extensive works that leveraged deep reinforcement learning to solve the IM problem. Existing state-of-the-art solutions \cite{li2019disco, tian2020deep, manchanda2020gcomb, li2022piano, chen2022touplegdd} follow a similar paradigm: learning latent embeddings of nodes or networks, and taking the current node embedding as the state of the agent in order to choose the next seed node as action, where the reward is its marginal influence gain.
    Other than reinforcement learning-based IM methods, there also exist methods \cite{kumar2022influence, kamarthi2019influence, panagopoulos2020multi} that solely leverage graph neural networks to encode social influence into node embedding and guide the node selection process. The crux of current learning-based IM methods is also obvious, the model complexity and adaptivity of learning-based IM methods are still not comparable to traditional methods. Particularly, current ML-based algorithms can neither handle the diversified diffusion patterns nor guarantee the quality of the solution as well as the model scalability.

    \subsection{Graph Neural Network}
    Graph Neural Networks (GNNs) \cite{wu2020comprehensive} are a class of deep learning methods designed to perform inference on data described by graphs. The general paradigm of GNNs alternates between node feature transformation and neighbor nodes' information aggregation. For a $K$-layer GNN, a node aggregates information within $K$-hop neighbors. Specifically, the $k$-th layer transformation is:
    \begin{align}
        a^{k} &= \mathcal{A}^k(h^{k-1};\theta^k),
         h^{k} = \mathcal{C}^k(a^{k};\theta^k), \forall 1\leq k \leq K.
         \label{eq:GNN}
    \end{align}
    where $a^{k}$ is an aggregated feature, and $h^{k}$ is the $k$-th layer node feature. The flexibility of aggregation function $\mathcal{A}(\cdot)$ and combine function $\mathcal{C}(\cdot)$ functions induces different GNN models~\cite{velivckovic2017graph, kipf2016semi, xu2018powerful,wang2022toward}. The high-level representations of nodes or graphs are utilized for different tasks. GNNs have been applied in various learning tasks such as information diffusion estimation~\cite{chamberlain2021grand, xia2021deepis, ko2020monstor}, graph source localization ~\cite{wang2022invertible,ling2022source}, deep graph generation \cite{ling2021deep,ling2023stgen,ling2023motif}, and graph analogical reasoning \cite{ling2022deepgar}. In this work, we leverage GNN to characterize the underlying diffusion pattern and construct an end-to-end model for estimating the influence.
    
\section{Problem Formulation}\label{sec:prob}
    Given a graph $G=\{V, E\}$, the problem of IM aims to maximize the number of influenced nodes in $G$ by selecting an optimal seed node set $\mathbf{x} \subseteq V$. Particularly, the evaluation of IM relies on an influence diffusion model parametrized by $\theta$: $\mathbf{y}=M(\mathbf{x}, G; \theta)$, where $\theta$ can be the set of infection probability on each node if $M(\cdot)$ is an independent cascade model or the set of parameters in the aggregation/combine functions if $M(\cdot)$ is GNN-based. We denote $\mathbf{x}\in \{0,1\}^{|V|}$ as the vector representation of the source node set, where the $i$-th element $x_i=1, x_i \in \mathbf{x}$ if $v_i\in \mathbf{x}$ and $x_i=0$ otherwise. 
    The output $y \in \mathbb{R}_+$ measures the total number of infected nodes~\cite{li2018influence}. Based on the formalization of the influence spread, the IM problem is defined as follows:
    \begin{dfn}[Influence Maximization]
    The generic \emph{IM} problem requires selecting a set of $k$ users from $V$ as the seed set to maximize the influence spread:
    \begin{equation}\label{eq: dfn}
        \Tilde{\mathbf{x}} = \argmax\nolimits_{|\mathbf{x}|\le k}  M(\mathbf{x}, G; \theta),
    \end{equation}
    where $\Tilde{\mathbf{x}}$ is the optimal seed node set that can produce a maximal influence spread in $G$.
    \end{dfn}
    Intuitively, selecting $\Tilde{\mathbf{x}}$ heavily depends on the underlying diffusion process. We have witnessed lots of works that develop algorithms with GNNs and reinforcement learning to tackle the problem. However, the expressiveness and generalization capability of existing learning-based IM frameworks is still limited due to the following challenges. 

    \noindent\textbf{Challenges.} Firstly, most existing learning-based IM frameworks calculate the latent node embedding for selecting high-influential ones. However, their objective functions require iteratively updating the latent embeddings for each node at each action/optimization step no matter whether they are included in the current $\mathbf{x}$. This poses a severe scalability problem if we are dealing with million-scale networks. Secondly, even though we leverage deep node/network embedding and various reward functions to guide the node selection process, existing frameworks are still tailored for specific diffusion models (e.g., they model $M(\cdot)$ as explicit IC and LT model).  However, these simple diffusion models cannot meet the needs of real applications. Moreover, to ease the computational overhead of the \#P-hard influence estimation, learning-based IM methods rely on techniques from traditional methods, such as proxy-based and sampling-based estimation way, which makes scalability and generalization even worse. Lastly,  there are plenty of node-centrality-constrained IM variants. For example, other than regulating the budget of the seed nodes, we may also need to regulate the total cost of selecting seed nodes. Learning-based IM solutions have different objective functions designed according to various application scenarios, and they do not have a well-defined scheme for all of the node-centrality-related constraints.    
    
    \section{DeepIM}\label{sec:model}
    In this section, we propose the DeepIM framework to ease the computational overhead of the learning-based IM methods and automatically identify the underlying diffusion patterns. The framework can be divided into two phases: the \emph{learning phase} is leveraged to characterize the probability of the observed seed set and model the underlying information propagation distribution, and the \emph{inference phase} is employed to optimize the selection of seeds in continuous space to maximize the influence spread.

    \begin{figure*}[!t]
 		\centerline{\includegraphics[width=\textwidth]{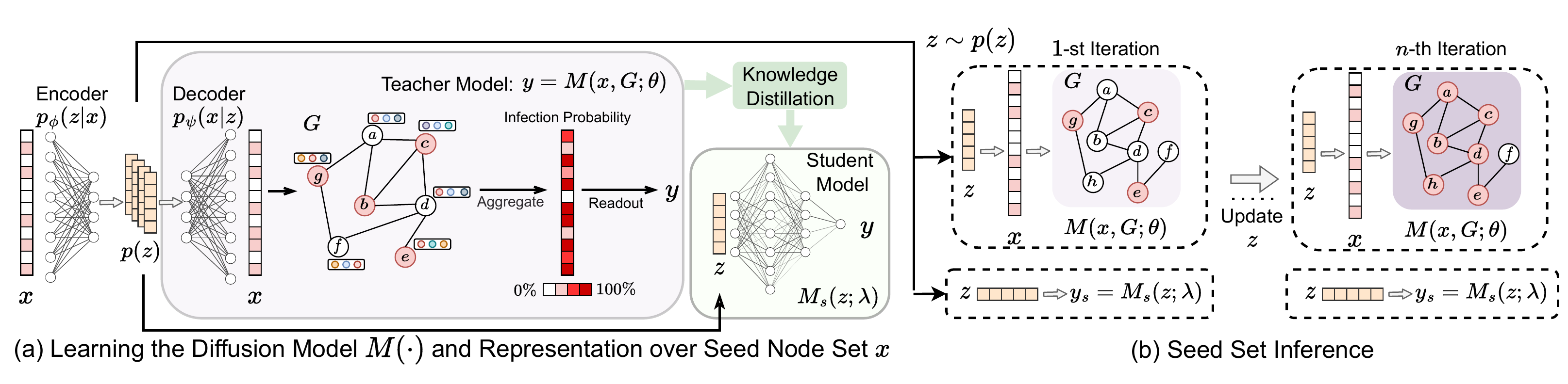}}
 		\vspace{-5mm}
 		\caption{DeepIM  consists of two parts. a) we leverage the autoencoder to learn and compress the latent distribution of seed node sets into lower dimension $p(z)$. The lower dimensional $p(z)$ is then leveraged to learn an end-to-end and monotonic diffusion model $M(\mathbf{x},G;\theta)$ for accurately predicting the spread. In addition, we employ a knowledge distillation module to train a lightweight student model to retain efficiency in predicting the influence spread. b) the seed set inference scheme iteratively optimizes the proposed objective function by updating the latent variable $z$, initially sampled from the learned $p(z)$, to maximize the influence spread.}
 		\label{fig: model}
 		\vspace{-5mm}
 	\end{figure*}

    \subsection{Learning Representation of Seed Set}
    To build an effective and efficient objective function, we propose to characterize the probability of the seed node set $p(\mathbf{x})$ over $\mathbf{x}$ given the graph $G$ since learning $p(\mathbf{x})$ can help depict the seed set's underlying nature. However, learning such probability is not a trivial task because different nodes are inter-connected within each seed set and highly correlated based on the topology of $G$. These connections make the relationship between nodes very complex and hard to decipher than other similar combinatorial problems.
    
    \noindent\textbf{Learning Probability over Seed Nodes.} Instead of directly modeling the highly-intractable probability $p(\mathbf{x})$, we introduce an unobserved latent variable $z$ to represent $\mathbf{x}$ and define a conditional distribution $p(\mathbf{x}\vert z)$ to quantify the likelihood. These latent variables have much lower dimensions than the observed sub-optimal seed sets, which can yield a compressed representation. Particularly, we marginalize over the latent variables to obtain $p(\mathbf{x}) = \int p(\mathbf{x}, z) \,dz = \int p(\mathbf{x}\vert z)p(z) \,dz.$
    The posterior likelihood $p(z\vert \mathbf{x}) = p(\mathbf{x}|z)\cdot p(z)/p(\mathbf{x})$ allows us to infer $z$ given the observed seed sets $\mathbf{x}$. In this work, we adopt autoencoder to generatively infer the posterior, where both encoder $f_{\phi}$ (parameterized by $\phi$) and decoder $f_{\psi}$ (parameterized by $\psi$) are used to characterize the likelihood of both posterior and conditional distribution, respectively. The objective of the autoencoder is to maximize the joint likelihood:
    \begin{align}\label{eq: representation_x}
        \max\nolimits_{\phi, \psi} \:\mathbb{E} \big[p_{\psi}(\mathbf{x}|z) \cdot p_{\phi}(z|\mathbf{x})\big].
    \end{align}
    
    \noindent\textbf{Learning the End-to-end Diffusion Model.}
    Once we have learned the latent distribution of seed nodes $p(\mathbf{x})$, the next step is to update the seed node set $\mathbf{x}$ in order to increase the marginal gain of the influence spread. Current learning-based IM solutions still assume the computation of the influence spread (i.e., $M(\mathbf{x},G;\theta)$) relies on prescribed mathematical models. However, real-world information diffusion is complicated, and it is not easy to determine the most suitable diffusion model in practice. A chosen diffusion model may be misspecified compared to real-world data and lead to large model bias. In addition, the diffusion network structure can also be hidden from us, so we need to learn not only the parameters in the diffusion model but also the diffusion network structure \cite{du2014influence}.

    In this work, we design a GNN-based diffusion model $M(\cdot)$ for accurate modeling of the relationship between $\mathbf{x}$ and $y$ with considering the overall graph topology. The output of a GNN-based diffusion function $M(\cdot)$ is composed of two functions $M = g_r\circ g_u(\mathbf{x},G; \theta)$: 1) $\tau= g_u(\mathbf{x},G; \theta)$, where $g_u(\cdot)$ is a GNN-based aggregation function and $\tau\in [0,1]^{|V|}$ is an intermediate output after aggregating multi-hop neighborhood information. $\tau$ denotes the \textit{infection probability} of each node; and 2) $y=g_r(\tau; \xi), y\in \mathbb{R}_+$ denotes the final information spread, where $g_r(\cdot)$ is a normalization function (e.g., $l$-1 norm) and $\xi$ is the threshold to transform the probability into discrete value. The GNN-based  $M(\cdot)$ is visualized in Fig. \ref{fig: model} (a). 
    \begin{dfn}[\textbf{Score Monotonicity and Infection Monotonicity}]        
        Given a GNN-based diffusion model $ M(\cdot): 2^{\vert V\vert}  \rightarrow \mathbb{R}_+$ and any two subsets $S, T \subseteq V$, $M(\cdot)$ is score monotonic if $x_S\preceq x_T$ (i.e. $S\subseteq T$) implies $M(\mathbf{x}_S,G; \theta) \le M(\mathbf{x}_T,G;\theta)$, where $\mathbf{x}_S, \mathbf{x}_T\in\{0,1\}^{\vert V\vert}$ are vector representations of seed sets $S$ and $T$, respectively. $M(\cdot)$ is infection monotonic if $x_S\preceq x_T$ (i.e. $S\subseteq T$) implies $\tau_S \preceq \tau_T$, where $\tau_S,\tau_T\in[0,1]^{\vert V\vert}$ denote the infection probability of seed sets $S$ and $T$, respectively.  
    \end{dfn}
    Monotonicity is a natural property for us in modeling the overall diffusion network structure. A monotonic diffusion model indicates the spread of influence would continue to increase. Intuitively, if we select a larger community $\mathbf{x}'$ as the seed set, the larger $\mathbf{x}'$ would intrinsically infect no less nodes in the whole network than a smaller seed set $\mathbf{x}$ if $\mathbf{x} \preceq \mathbf{x}'$. Ensuring the property of both monotonicities allows us to better characterize the underlying diffusion network structure and mimic the real-world diffusion pattern \cite{dolhansky2016deep}. Hence, we add constraints to make the GNN-based diffusion model $M(\mathbf{x}, G; \theta)$ monotonic during the influence spread estimation.
        
    \begin{thm}[\textbf{Monotonicity of GNN Models}]
	\label{theorem: submodular}
	For any GNN-based $M(\mathbf{x}, G;\theta)=g_r\circ g_u(\mathbf{x},G; \theta)$, where $g_u(\mathbf{x},G; \theta)$ is formulated by Eq. \eqref{eq:GNN}, $M$ is score and infection monotonic if $\mathcal{A}^k$ and $C^k, k\in [1,K]$, are non-decreasing in Eq. \eqref{eq:GNN}, and $g_r$ is also non-decreasing.
	\end{thm}
	We further illustrate that the well-known Graph ATtention network (GAT) can be score and infection monotonic under the constraint we claimed in Theorem \ref{theorem: submodular}. Note that we follow the standard network structures of GAT as introduced in the original paper.
	\begin{cor}[\textbf{Montonicity of GAT}]\label{theorem: gcn_gat} $M$ is score and infection monotonic when $g_u$ is GAT if $\theta^k\geq 0$ in Eq. \eqref{eq:GNN} and $g_r$ is also non-decreasing.
	\end{cor}
    Due to the space limitation, the proof of Theorem \ref{theorem: submodular} and Corollary \ref{theorem: gcn_gat} are provided in Appendix \ref{sec: proofs}. According to Theorem \ref{theorem: submodular} and Corollary \ref{theorem: gcn_gat}, the GNN-based $M(\mathbf{x}, G; \theta)$ has the theoretical guarantee to retain monotonicity, and the objective of learning the GNN-based $M(\mathbf{x}, G; \theta)$ is given as maximizing the following probability with a constraint:
    \begin{align}\label{eq: representation_y}
        \max\nolimits_{\theta}\mathbb{E}\big[p_{\theta}(y|\mathbf{x}, G)\big], \quad \text{s.t. } \theta \ge 0.
    \end{align}

    \noindent\textbf{Knowledge Distillation for Diffusion Estimation Efficiency.} We have learnt the deep representation of seed nodes and an end-to-end diffusion model with a monotonicity guarantee. However, we empirically find the calculation of influence spread $M(\mathbf{x},G;\theta)$ involves three steps: 1) decoding a node vector $\mathbf{x}$ from the learned posterior $p(\mathbf{x}|z)$; and 2) executing the GNN-based diffusion model $M(\mathbf{x},G;\theta)$ under the graph $G$; and 3) normalizing the probabilistic output $\tau$ from $M(\mathbf{x},G;\theta)$ to actual influence spread $y$. Even though the prediction results are accurate, the computational overhead is still a burden when dealing with million-scale networks. Inspired by recent research on knowledge distillation, we propose to leverage a small yet powerful student model supervised by $M(\mathbf{x},G;\theta)$ to attain efficiency. Specifically, the student model $M_s(z; \lambda)$ is a lightweight neural network parametrized by $\lambda$ that directly takes the latent variable $z$ sampled from the learned $p(z)$ as input. $M_s(z; \lambda)$ directly returns the estimated influence spread $y_s$ as output. The distillation loss between the $y=M(\mathbf{x},G;\theta)$ (teacher model) and $y_s = M_s(z;\lambda)$ can be as simple as $\norm{y-y_s}_2^2$.
        
    \noindent\textbf{End-to-end Learning Objective.} Finally, in order to bridge the representation learning and the learning of the diffusion model, we propose a unified objective function in an end-to-end manner by putting together Eq. \eqref{eq: representation_x} and \eqref{eq: representation_y} as:
    \begin{align}\label{eq: learning_objective_1}
    \mathcal{L}_{\text{train}}=\max\nolimits_{\theta, \lambda, \psi, \phi} &\mathbb{E}\big[ p_{\theta}(y|\mathbf{x}, G) \cdot p_{\lambda}(y_s| z)\\ &\hspace{5mm}\cdot p_{\psi}( \mathbf{x}|z)\cdot p_{\phi}(z|\mathbf{x})\big],\: \text{s.t. } \theta \ge 0.\nonumber
    \end{align}
    However, optimizing the expectation of joint probabilities could be computationally difficult. We instead derive the negative $\log$ term of Eq. \eqref{eq: learning_objective_1} and derive its lower bound as the final learning objective according to Jensen's inequality:
    \begin{align}\label{eq: learning_objective_2}
    \mathcal{L}_{\text{train}}&=\min\nolimits_{\theta, \lambda, \psi, \phi}-\log \Big[\mathbb{E}\big[ p_{\theta}(y|\mathbf{x}, G) \cdot p_{\lambda}(y_s| z) \nonumber\\ &\hspace{10mm}\cdot p_{\psi}(\mathbf{x}|z)\cdot p_{\phi}(z|\mathbf{x})\big]\Big],\: \text{s.t. } \theta \ge 0.\nonumber\\
    \hspace{-6mm}&\ge
        \min\nolimits_{\theta, \lambda, \psi, \phi} \mathbb{E}\Big[ -\log\big[p_{\theta}(y|\mathbf{x}, G)\cdot p_{\lambda}(y_s| z)\nonumber\\ &\hspace{10mm}\cdot p_{\psi}(\mathbf{x}|z)\cdot (p_{\phi}(z|\mathbf{x})\big]\Big], \text{s.t. } \theta \ge 0.\nonumber\\
        \hspace{-6mm}&=
        \min\nolimits_{\theta, \lambda, \psi, \phi} \mathbb{E}\Big[ -\log\big[p_{\theta}(y|\mathbf{x}, G)\big] -\log\big[p_{\lambda}(y_s| z)\big]\nonumber\\ &\hspace{10mm}-\log\big[p_{\psi}(\mathbf{x}|z)\cdot (p_{\phi}(z|\mathbf{x})\big]\Big], \text{s.t. } \theta \ge 0.
    \end{align}
    The overall objective consists of minimizing the empirical error $-\log [p_{\theta}(y|\mathbf{x}, G)]$ of the prediction of $y$ with the reconstructed $\mathbf{x}$ as input and minimizing the reconstruction error. In addition, we minimize the distillation loss $-\log [p_{\lambda}(y_s| z)]$ to train the student model along with the overall training process. The overall framework for the training of end-to-end diffusion models and the autoencoder for learning the seed set distribution is visualized in Fig. \ref{fig: model} (a).

    \subsection{Seed Node Set Inference}
    To infer the high-influential seed node set in the testing domain, we leverage the latent distribution $p(\mathbf{x})$ of the seed node set and the end-to-end diffusion model $M(\cdot)$ jointly from Eq. \eqref{eq: learning_objective_2}. Firstly, if the autoencoder is well trained and can retain both \textit{continuity} (i.e., two close points in the latent space should not give two completely different contents once decoded) and \textit{completeness} (i.e., for a chosen distribution, a point sampled from the latent space should give “meaningful” content once decoded), the autoencoder in Eq. \eqref{eq: representation_x} can generate contents by exploiting the latent feature space $p(z)$ learned from all the examples it was trained from, i.e., $p(\mathbf{x})$. Therefore, we propose to alternatively search the optimal seed node set $\Tilde x$ in the lower-dimensional and less-noisy latent space $p(z)$. The following corollary demonstrates it is equivalent to estimating the influence spread with the latent variable $z$ rather than high-dimensional $\mathbf{x}$ if the autoencoder retains both continuity and completeness.
    \begin{cor}[\textbf{Influence Estimation Consistency}]\label{theorem: consistency}
	For any \\$M(f_{\psi}(z^{(i)}), G; \theta)>M(f_{\psi}(z^{(j)}), G; \theta)$, we have $M(x^{(i)}, G; \theta) > M(x^{(j)}, G; \theta)$.
	\end{cor}
    
    The proof of Corollary \ref{theorem: consistency} can be found in Appendix. According to the corollary, we could find the optimal seed set that can generate the maximal influence by optimizing $z$ in the following joint probability: $\max_z \mathbb{E}\big[ p_{\theta}(y|\mathbf{x}, G)\cdot p_{\psi}(\mathbf{x}|z)\big]$. 
    
    \noindent \textbf{Adaptation to Different IM Variants with Node Centrality Constraints.} Since the introduction of IM in \cite{kempe2003maximizing}, IM was studied under various budget constrained settings on nodes in recent years. To enhance the adaptivity of DeepIM, we design a unified constraint that allows inferring seed sets under various budgets on individual nodes. Specifically, the objective $\mathcal{L}_{\text{pred}}$ is given as:
    \begin{align}\label{eq: infer}
        \mathcal{L}_{\text{pred}} &= \max_z \mathbb{E}\big[ p_{\theta}(y|\mathbf{x}, G)\cdot p_{\psi}(\mathbf{x}|z)\big], \nonumber\\ &\text{s.t.} \: \sum\nolimits^{|V|}_{i=0} \mathcal{F}(v_i, G)\cdot x_i \le k,
    \end{align}
    where $\sum\nolimits^{|V|}_{i=0} \mathcal{F}(v_i, G)\cdot x_i$ is a generalized budget constraint applied on individual nodes, and $k$ is the actual budget. For the vanilla IM problem that only requires selecting a given number of seed nodes, $\sum\nolimits^{|V|}_{i=0} \mathcal{F}(v_i, G)\cdot x_i$ can be derived as $\norm{x \cdot \mathbf{1}}_1$, where the $\mathbf{1} \in \{1\}^{N\times 1}$ is an all-one vector indicating the price of selecting each node are the same. In addition, for node degree constrained IM problems~\cite{leskovec2007cost,nguyen2017billion}, $\sum\nolimits^{|V|}_{i=0} \mathcal{F}(v_i, G)\cdot x_i$ can be derived as $\norm{x \cdot A}_1$, where $A \in \{0, 1\}^{N\times N}$ is the adjacency matrix of the network $G$, and $\norm{\mathbf{x} \cdot A^i}_1 \le k$ represents the $l1$-norm of the total seed node degree is bounded by a budget $k$. The budget constraint $\mathcal{F}(v_i, G)\cdot x_i \le k$ can also be easily designed, combined, and adapted to solve the IM variants with even non-uniform prices on each node. With the proposed flexible constraint, the challenge of IM methods' adaptability can be partially addressed. 
            
    \begin{algorithm}[t]
    \caption{DeepIM Prediction Framework}\label{algo: inference_framework}
    \begin{algorithmic}[1]
    \renewcommand{\algorithmicrequire}{\textbf{Input:}}
    \renewcommand{\algorithmicensure}{\textbf{Require:}}
    \REQUIRE $\mathcal{L}_{\text{pred}}$; $f_{\psi}(\cdot)$; $\Phi(\cdot)$; number of training instances $N$; the number of iteration $\eta$; learning rate $\alpha$.\\
    \STATE $z = 1/N \sum_{i=0}^{N}f_{\psi}(\mathbf{x})$ \COMMENT{$\mathbf{x}$ sampled from training set.}
    \FOR {$i=0, ..., \eta$}
        \STATE $\mathbf{x} \leftarrow f_{\psi}(z)$ \COMMENT{seed set $\mathbf{x}$.}
        \STATE $\mathbf{x} \leftarrow \Phi(\mathbf{x})$ \COMMENT{Regularize $\mathbf{x}$ into valid regions.}
        \STATE $z \leftarrow z - \alpha\cdot \nabla \mathcal{L}_{\text{pred}}(\mathbf{x}, z)$
    \ENDFOR\\
    \STATE $\Tilde{\mathbf{x}} \leftarrow \Phi(f_{\psi}(z))$ 
    \end{algorithmic}
    \end{algorithm}
    
    \noindent\textbf{Implementation Details of the Seed Set Inference.} We visualize our inference procedure in Fig. \ref{fig: model} (b). Specifically, the inference framework first samples a latent variable $z$ from the learned latent distribution $p(z)$. The latent variable $z$ is iteratively optimized according to the inference objective function Eq. \eqref{eq: infer} to attain a larger marginal gain (influence spread). Note that the learning-based diffusion model $p_{\theta}(y|\mathbf{x},G)$ can be switched between the student diffusion model $M_{s}(z; \lambda)$ and the GNN-based diffusion model $M(\mathbf{x}, G; \theta)$ to achieve either efficiency or efficacy. In addition, the constrained objective function Eq. \eqref{eq: infer} cannot be computed directly so that we provide a practical version of the inference objective function: since the diffused observation $y$ fits the Gaussian distribution and the seed set $\mathbf{x}$ fits the Bernoulli distribution, we can simplify Eq. \eqref{eq: infer} as:
    \begin{align}\label{eq: infer_2}
        \mathcal{L}_{\text{pred}} = &\min_z \Big[- \log \big[\prod\nolimits_{i=0}^{|V|}f_{\psi}(z_i)^{x_i}(1-f_{\psi}(z_i)^{1-x_i}\big]\nonumber\\ &\hspace{-6mm}+\big\rVert \Tilde y- y\big\rVert_2^2 \Big] \:\text{s.t. }  \sum\nolimits^{|V|}_{i=0} \mathcal{F}(v_i, G)\cdot x_i \le k,
    \end{align}
    where the $\Tilde y$ is given as the optimal influence spread (i.e., $\Tilde y = |V|$), and the full derivation of the above equation is provided in Appendix. Furthermore, we utilize the Projected Gradient Descent and propose a regularization function $\Phi(\mathbf{x})$ to keep the predicted seed set $\mathbf{x}$ in a valid region in terms of different constraints. For example, $\Phi(\mathbf{x})$ can be defined as selecting $k$ nodes with the highest probabilities when the price of selecting each node is equal in Eq. \eqref{eq: infer}. $\Phi(\mathbf{x})$ can also be defined as cost-efficiently selecting the top-$k$ nodes from $\mathbf{x}/c(\mathbf{x})$, where $c(\mathbf{x})$ denotes the budget on one node (e.g., node degree). Finally, The optimization procedure is summarized in Algorithm \ref{algo: inference_framework}. Specifically, we firstly sample an initial latent variable $z$ on Line $1$. From Line $2$ - $6$, we iteratively solve the optimization problem proposed in Eq. \eqref{eq: infer} via gradient descent optimizer (e.g., Adam) while regularizing the predicted seed set in a valid region with $\Phi(\cdot)$. Fig. \ref{fig: model} (b) illustrates the overall process of the inference objective learning. We provide the derivation details of both Eq. \eqref{eq: learning_objective_2} and \eqref{eq: infer_2} in Appendix \ref{sec: proofs}.
    
    \begin{table}[t]
\resizebox{0.49\textwidth}{!}{%
\begin{tabular}{@{}cccccccc@{}}
\toprule
      & Digg      & Weibo       & \begin{tabular}[c]{@{}c@{}}Power \\ Grid\end{tabular} & \begin{tabular}[c]{@{}c@{}}Network\\ Science\end{tabular} & Cora-ML & Jazz  & Synthetic \\ \midrule
Nodes & 279,613   & 2,251,166   & 4,941                                                 & 1,565                                                     & 2,810   & 198   & 50,000    \\
Edges & 1,170,689 & 225,877,808 & 6,594                                                 & 13,532                                                    & 7,981   & 2,742 & 250,000   \\ \bottomrule
\end{tabular}}
\vspace{-4mm}
\caption{The Overview of Dataset}
\vspace{-7mm}
\label{tab: dataset}
\end{table}

    \section{Experiment}\label{sec:exp}
	In this section, we compare the performance of our proposed DeepIM framework across six real networks in maximizing the influence under various settings, following a case study to qualitatively demonstrate the performance of DeepIM. Due to the space limit, more details of the experiment setup, hyperparameter settings, dataset description, and comparison methods can be found in Appendix.

	\subsection{Experiment Setup}
	Our primary purpose is to evaluate the expected influence spread as defined in Eq. \eqref{eq: dfn} under various IM application scenarios. Since DeepIM can be easily adapted to different diffusion patterns, we choose two representative models that are commonly used in the IM problem, i.e., LT and IC model. In addition, we also evaluate the IM problem under the susceptible-infectious-susceptible (SIS) epidemic model~\cite{kermack1927contribution}, where the major difference is activated nodes can be de-activated in SIS. Due to the space limit, the experiment of the non-progressive diffusion model can be found in Appendix. 
 
	\noindent\textbf{Data.} The proposed DeepIM is compared with other approaches over six real-world datasets, including Cora-ML, Network Science, Power Grid, Jazz, Digg, and Weibo. We also adopt a synthetic dataset that is a random graph with $50,000$ nodes generated by Erdos-Renyi algorithm~\cite{erdHos1960evolution}. The statistics of the data are shown in Table \ref{tab: dataset}. We randomly sample seed node set $\mathbf{x}$, and the seed size is proportional to $|V|$ of each network. We then use IC, LT, and SIS models to compute the final influence spread $y$. The $\{(\mathbf{x}, y)\}$ set then serves as the training set of our algorithm.

 	\begin{table*}[t]
    \centering
    \resizebox{\textwidth}{!}{%
    \begin{tabular}{@{}c|cccc|cccc|cccc|cccc|cccc|cccc|cccc@{}}
    \toprule
        & \multicolumn{4}{c|}{Cora-ML}         & \multicolumn{4}{c|}{Network Science}      & \multicolumn{4}{c|}{Power Grid}   &  \multicolumn{4}{c|}{Jazz} &
        \multicolumn{4}{c|}{Synthetic} &
        \multicolumn{4}{c|}{Digg} &
        \multicolumn{4}{c}{Weibo}\\ \midrule
    Methods & 1\%     & 5\%     & 10\%     & 20\%    & 1\%     & 5\%     & 10\%     & 20\%    & 1\%     & 5\%     & 10\%     & 20\%    & 1\%     & 5\%     & 10\%     & 20\%   & 1\%     & 5\%     & 10\%     & 20\%   & 1\%     & 5\%     & 10\%     & 20\%   & 1\%     & 5\%     & 10\%     & 20\%\\ \midrule
    IMM            & 8.1          & 26.2          & 37.3          & 50.2          & 5.2           & 16.8          & 27.0          & 45.7          & 4.3          & 17.4           & 31.5          & 51.1          & 2.6          & 20.1          & 31.4          & 42.8           & 9.2          & 26.2          & 36.3          & 51.6          
    & 7.4          & 18.4          & 32.8          & 49.6
    & 9.5          & 23.8          & 36.4          & 50.5\\
    OPIM           & 13.4          & 26.9          & 37.4          & 50.9          & 6.6          & 19.4          & 28.9          & 48.6          & 5.7          & 17.7           & 29.7          & 50.1          & 2.4          & 20.1          & 34.4          & 46.8           & 9.6          & 25.3          & 36.6          & 51.7          
    & 7.6          & 18.5          & 32.9          & 48.9
    & 9.7          & 23.7          & 36.6          & 50.3\\
    SubSIM         & 10.1          & 25.7          & 36.8          & 51.1          & 4.8           & 15.4          & 27.9          & 44.8          & 4.6          & 19.2           & 31.7          & 50.2          & 3.6          & 18.8          & 37.6          & 44.7           & 9.5          & 26.7          & 36.5          & 51.5         
    & 7.5          & 18.9          & 33.3          & 49.4
    & 9.3          & 23.1          & 36.5          & 50.6\\ \midrule
    OIM            & 8.9          & 27.6          & 38.0          & 51.3          & 4.2           & 16.7          & 26.5          & 48.2          & 5.7          & 17.5           & 31.9          & 50.8          & 2.0          & 18.5          & 36.3          & 42.2           &9.6           &26.2                  &36.7                  &51.3     & 7.8          & 18.2          & 33.1          & 49.6
    & -          & -          & -          & -\\\midrule
    IMINFECTOR         & 9.6          & 26.8          & 37.7          &50.6          & 5.4          & 17.9          & 27.8          & 47.6          & 5.4          & 18.2           & 31.6          & 50.9          & 3.6         & 19.7          & 37.5          & 45.9           & 9.1          & 26.2          & 36.1          & 51.5
    & 7.9          & 18.6          & 33.5         & 49.8
    & 9.4         & 23.5          & 36.9          & 50.3\\
     PIANO        & 9.8          & 25.2          & 37.4          & 51.1          & 4.7           & 16.3          & 27.1          & 47.2          & 5.3          & 18.1           & 31.7          & 50.2          & 2.2          & 19.2          & 36.6          & 43.2           &9.1           &26.4                  &36.2                  &51.6     & -          & -          & -          & -
    & -          & -          & -          & -\\
     ToupleGDD       & 10.6          & 27.5          & 38.5          & 51.5          & 6.3           & 17.8          & 28.3          & 50.5          & 5.4          & 19.3           & 31.6          & 51.3          & 3.3          & 20.4          & 37.2          & 45.7          &9.5           &26.8                  &37.1                  &51.4     & -          & -          & -          & -
    & -          & -          & -          & -\\\midrule
    DeepIM$_s$         & 13.6           & 27.7           & 38.5          & 51.8          & 6.9          & 19.1          & 29.3          & 50.5          & 5.9          & 20.2          & 31.7          & 51.5          & 3.8          & 21.4           & 38.9          & 47.1           & 10.2          & 26.8          & 37.5          & 51.8          & 7.9          & 18.8          & 33.7          & 50.3
    & 10.1          & 24.7          & 36.8          & 50.8\\
    \textbf{DeepIM} & \textbf{14.1}          & \textbf{28.1}          & \textbf{39.6}          & \textbf{52.4}          & \textbf{7.8}          & \textbf{20.9}          & \textbf{31.5}          & \textbf{51.2}          & \textbf{6.3}          & \textbf{21.0}          & \textbf{32.5}          & \textbf{52.4}          & \textbf{4.9}          & \textbf{23.3}          & \textbf{41.5}          & \textbf{49.9}           & \textbf{11.6} & \textbf{27.4}          & \textbf{38.7} & \textbf{52.1
}  & \textbf{8.4}          & \textbf{19.3}          & \textbf{34.2}          & \textbf{51.3}
& \textbf{11.2}          & \textbf{26.5}          & \textbf{37.9}          & \textbf{51.8}\\
    \bottomrule
    \end{tabular}
    }
    \vspace{-4mm}
    \caption{Performance comparison under IC diffusion pattern. $-$ indicates out-of-memory error. (Best is highlighted with bold.)}
    \vspace{-3mm}
    \label{tab: evaluation_ic}
    \end{table*}

    \begin{table*}[t]
    \centering
    \resizebox{\textwidth}{!}{%
    \begin{tabular}{@{}c|cccc|cccc|cccc|cccc|cccc|cccc|cccc@{}}
    \toprule
        & \multicolumn{4}{c|}{Cora-ML}         & \multicolumn{4}{c|}{Network Science}      & \multicolumn{4}{c|}{Power Grid}   &  \multicolumn{4}{c|}{Jazz} &
        \multicolumn{4}{c|}{Synthetic} &
        \multicolumn{4}{c|}{Digg} &
        \multicolumn{4}{c}{Weibo}\\ \midrule
    Methods & 1\%     & 5\%     & 10\%     & 20\%    & 1\%     & 5\%     & 10\%     & 20\%    & 1\%     & 5\%     & 10\%     & 20\%    & 1\%     & 5\%     & 10\%     & 20\%   & 1\%     & 5\%     & 10\%     & 20\%   & 1\%     & 5\%     & 10\%     & 20\%   & 1\%     & 5\%     & 10\%     & 20\%\\ \midrule
    IMM            & 1.7           & 34.8           & 52.2          & 66.4          & 2.5          & 11.9          & 18.1          & 33.6          & 4.6          & 19.9           & 31.7          & 56.9          & 1.4          & 5.7           & 13.4          & 24.5           & 1.1          & 5.2          & 13.1          & 66.9          & 2.4          & 10.8          & 37.4           & 55.6 
    & 1.6          & 6.7          & 19.3           & 45.2 \\
    OPIM           & 2.3           & 36.9          & 51.2          & 71.5          & 1.6          & 12.0          & 18.8          & 34.1          & 4.4          & 21.6          & 29.4          & 55.5          & 1.4          & 6.9           & 12.6          & 20.9           & 1.3          & 5.2          & 12.6          & 62.1          & 2.1          & 11.3         & 38.2           & 57.1 
    & 1.8          & 6.1          & 18.7           & 46.6 \\
    SubSIM         & 1.7           & 33.6           & 54.7          & 70.1          & 1.8          & 10.4          & 19.2          & 34.1          & 4.5          & 21.1          & 31.2          & 57.4          & 1.4          & 5.9           & 11.4          & 21.2           & 1.4          & 5.5          & 13.1          & 69.6          & 2.4          & 11.3          & 37.9           & 56.9 
    & 1.7          & 6.7          & 19.2           & 46.8\\ \midrule
    IMINFECTOR         & 2.1          & 33.9          & 51.3          & 70.6          & 2.1           & 11.8          & 18.7          & 34.5          & 4.2          & 21.3           & 31.6          & 56.2          & 1.4         & 6.2          & 13.5          & 22.8           & 1.3          & 5.2          & 12.9          & 67.4
    & 2.2          & 11.1          & 38.9         & 58.7
    & 1.8         & 6.4          & 18.6          & 47.5\\
    PIANO         & 2.1           & 33.5           & 53.3          & 69.8          & 2.1          & 11.3          & 19.1          & 33,9          & 4.3          & 21.3          & 31.4          & 57.1          & 1.1          & 6.2           & 12.1          & 22.4           & 1.2          & 5.2          & 12.9          & 67.4          & -          & -          & -          & -
    & -          & -          & -          & -\\
    ToubleGDD         & 2.3           & 36.2           & 54.5          & 70.9          & 2.8          & 12.4          & 19.8          & 34.6          & 4.8          & 21.9          & 32.6          & 58.1          & 1.4          & 6.5           & 12.9          & 23.6           & 1.3          & 5.5          & 13.4          & 70.2          & -          & -          & -          & -
    & -          & -          & -          & -\\\midrule
    DeepIM$_s$         & 10.7           & 65.6           & 75.1          & 85.2          & 3.5          & 14.6          & 23.8          & 37.8          & 5.1          & 22.9          & 40.3          & 65.1          & 1.4          & 6.5           & 14.2          & 85.3           & 1.5          & 6.0          & 14.2          & 90.3          & 3.1          & 13.3          & 39.2          & 67.9
    & 2.5          & 7.1          & 32.6          & 68.4\\
    \textbf{DeepIM} & \textbf{13.4} & \textbf{69.2} & \textbf{83.5} & \textbf{94.1} & \textbf{4.1} & \textbf{16.6} & \textbf{26.7} & \textbf{41.5} & \textbf{6.3} & \textbf{24.4} & \textbf{46.8} & \textbf{71.7} & \textbf{1.9} & \textbf{6.5} & \textbf{16.4} & \textbf{99.1} & \textbf{1.5} & \textbf{6.5} & \textbf{15.5} & \textbf{99.9}  & \textbf{3.5}          & \textbf{15.9}          & \textbf{41.3}           & \textbf{76.2}
    & \textbf{3.1}          & \textbf{7.6}          & \textbf{39.3}           & \textbf{72.4}\\
    \bottomrule
    \end{tabular}
    }
    \vspace{-4mm}
    \caption{Performance comparison under LT diffusion pattern. $-$ indicates out-of-memory error. (Best is highlighted with bold.)}
    \vspace{-6mm}
    \label{tab: evaluation_lt}
    \end{table*}

     \begin{figure*}[!t]
		\subfloat[Cora\_ML-IC]{\label{fig: coraml_budget_ic}
			\hspace{-3mm}\includegraphics[width=0.2\textwidth]{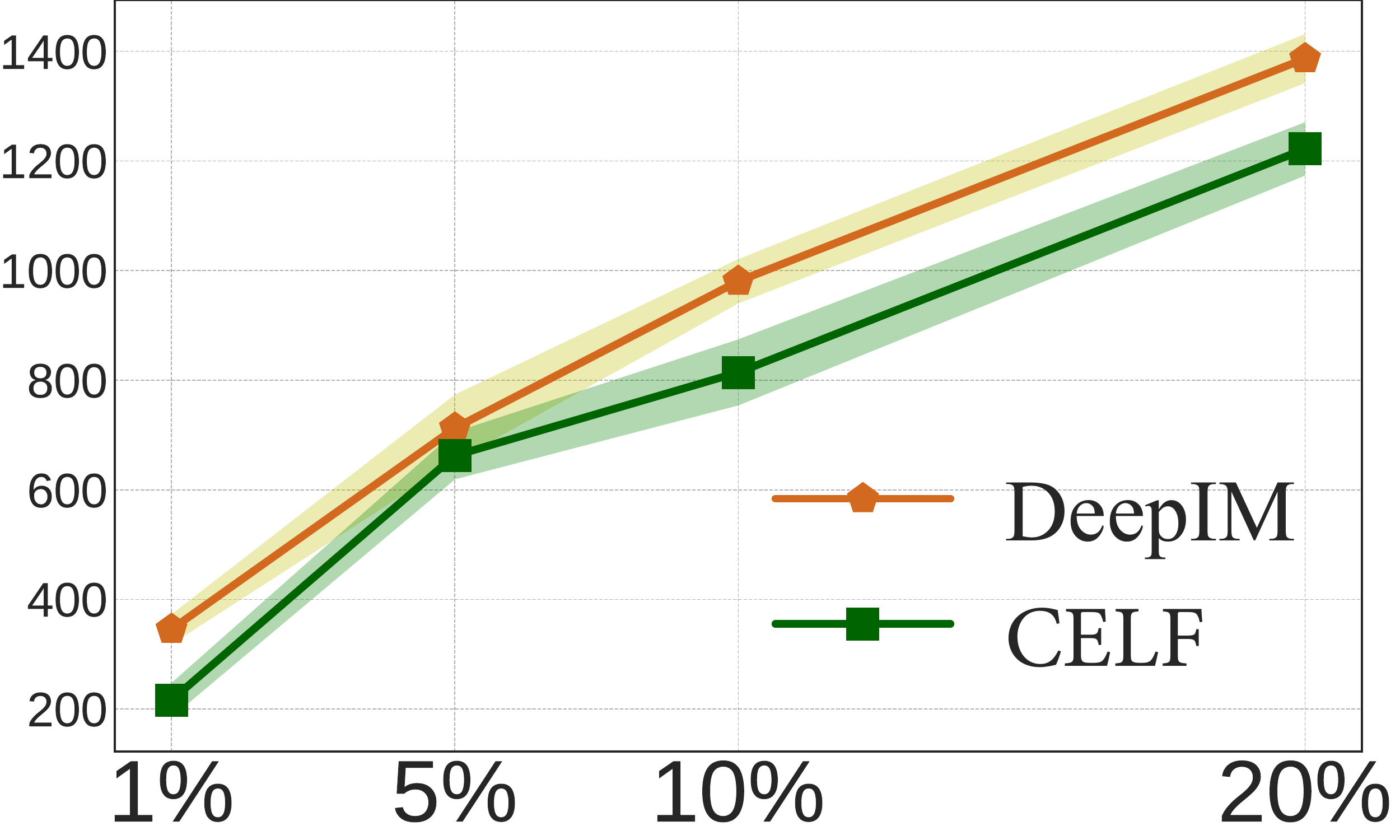}}
		\subfloat[Net Science-IC]{\label{fig: net_budget_ic}
			\includegraphics[width=0.2\textwidth]{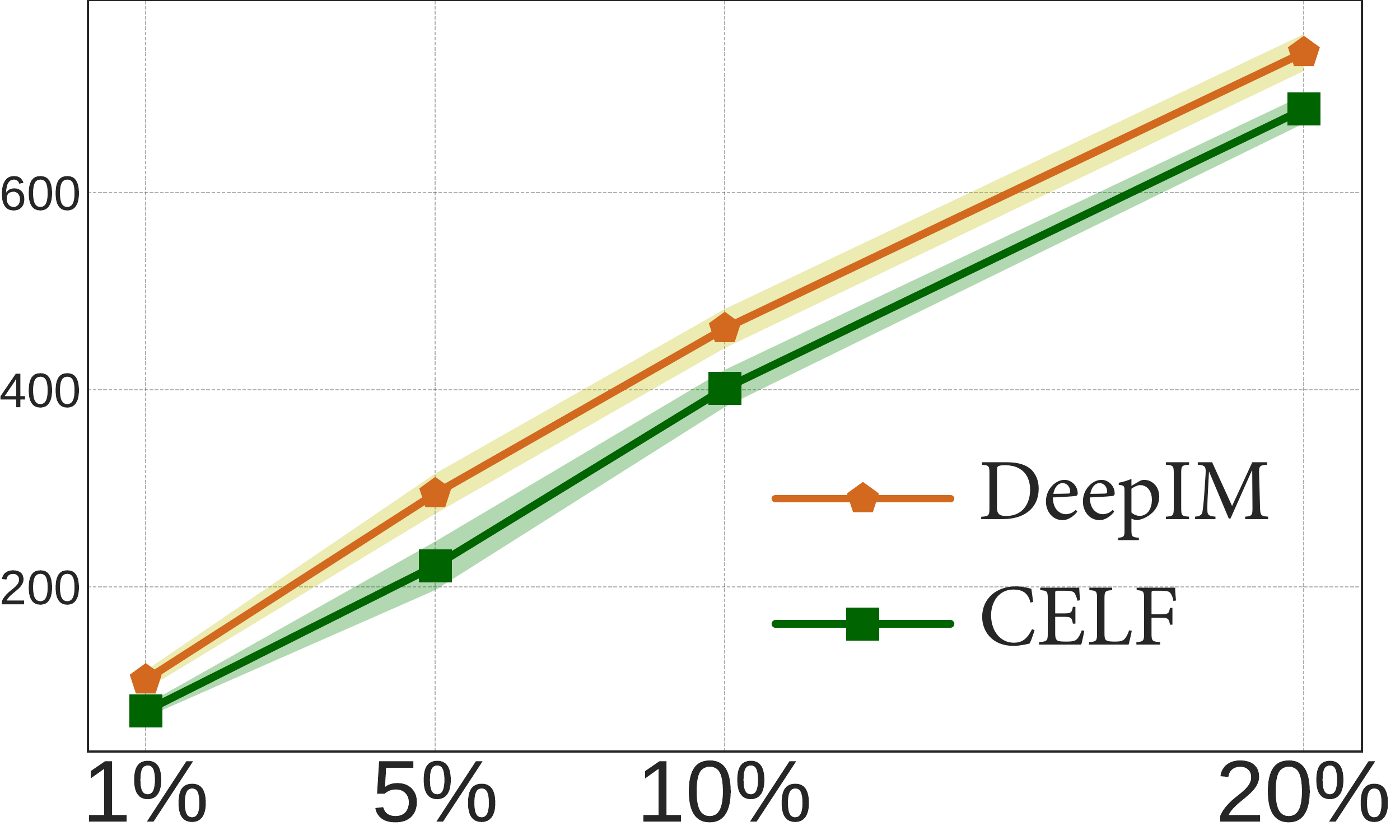}}
		\subfloat[Power Grid-IC]{\label{fig: power_budget_ic}
			\includegraphics[width=0.2\textwidth]{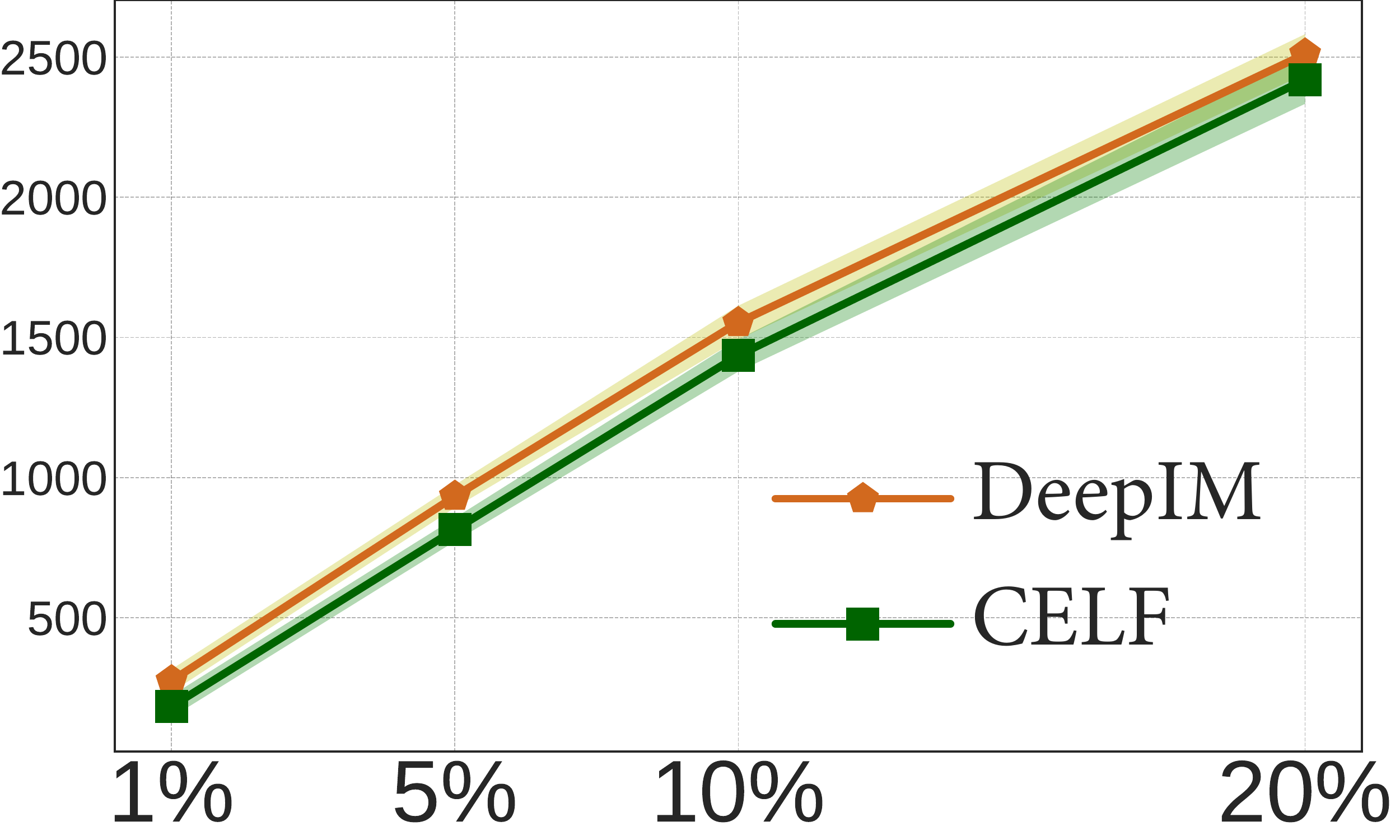}}
		\subfloat[Jazz-IC]{\label{fig: jazz_budget_ic}
			\includegraphics[width=0.2\textwidth]{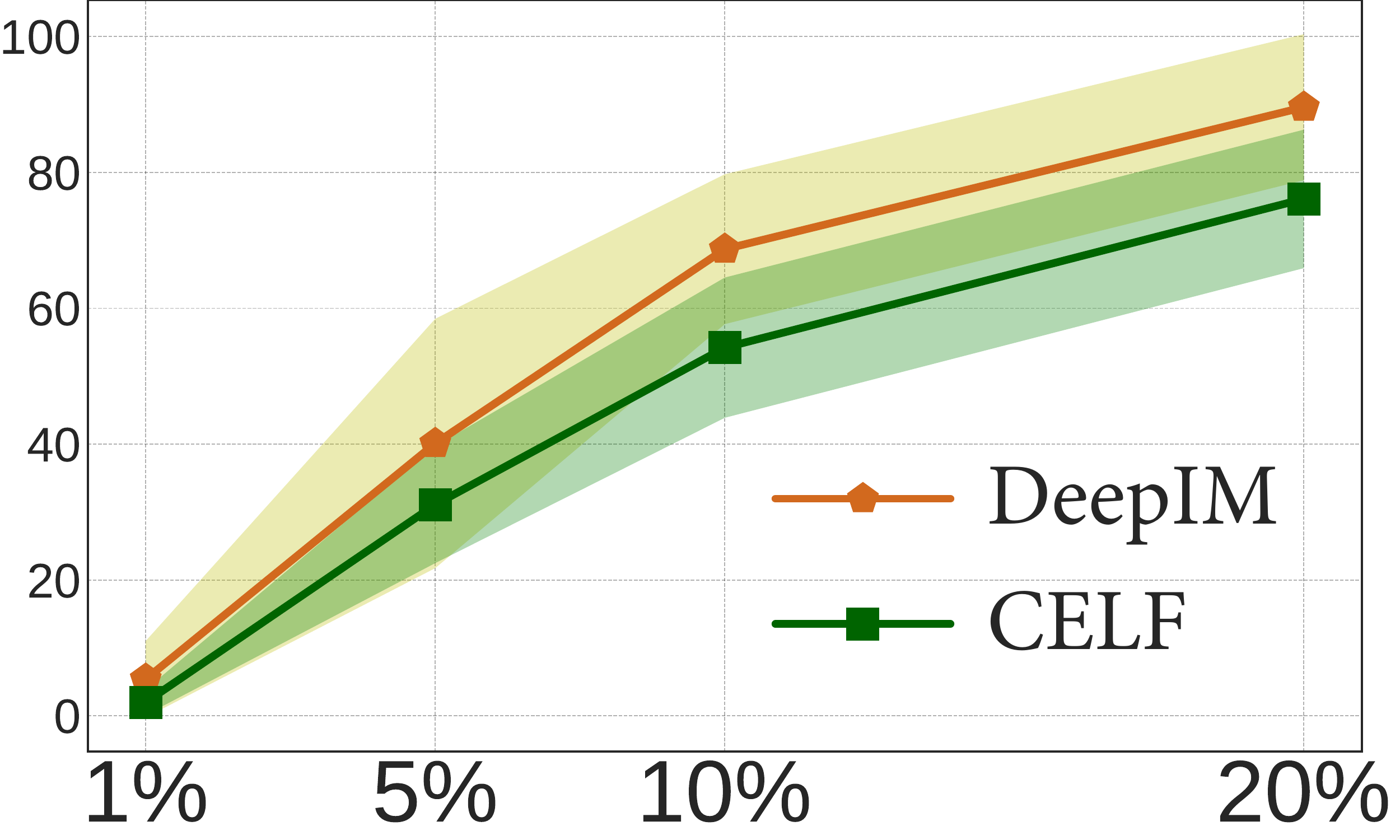}}
		\subfloat[Synthetic-IC]{\label{fig: syn_budget_ic}
			\includegraphics[width=0.2\textwidth]{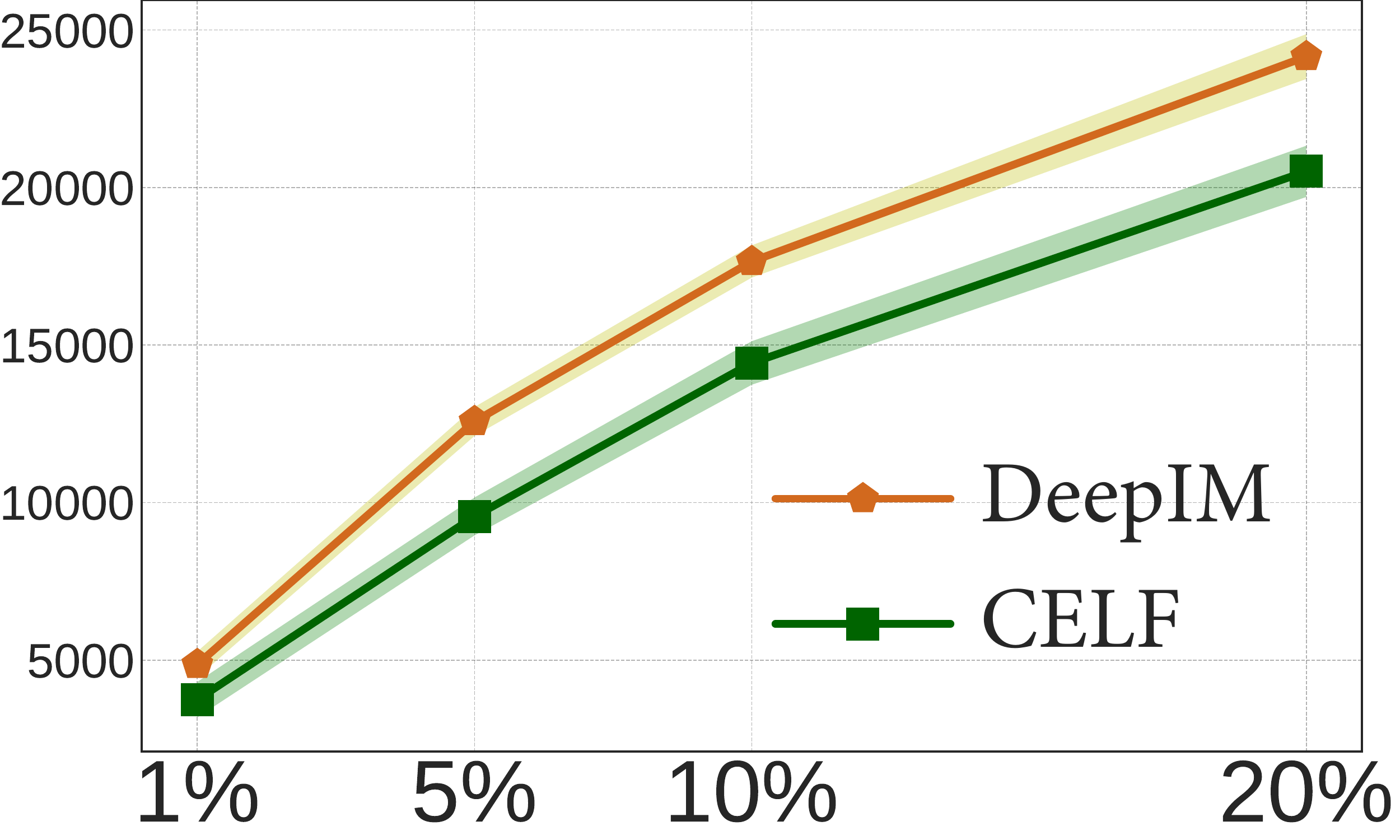}}
		\vspace{-3mm}
		\subfloat[Cora\_ML-LT]{\label{fig: coraml_budget_lt}
			\hspace{-3mm}\includegraphics[width=0.2\textwidth]{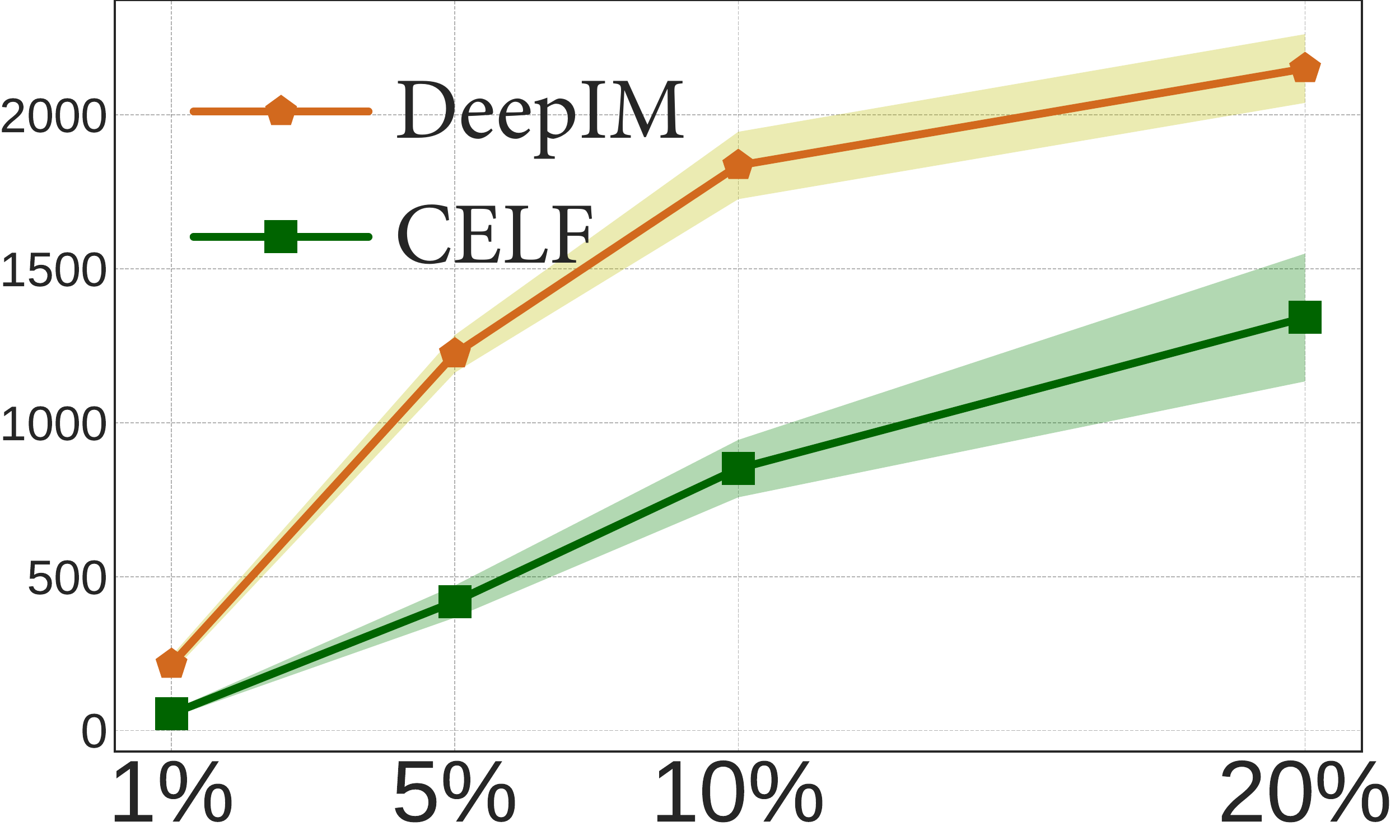}}
		\subfloat[Net Science-LT]{\label{fig: net_budget_lt}
			\includegraphics[width=0.2\textwidth]{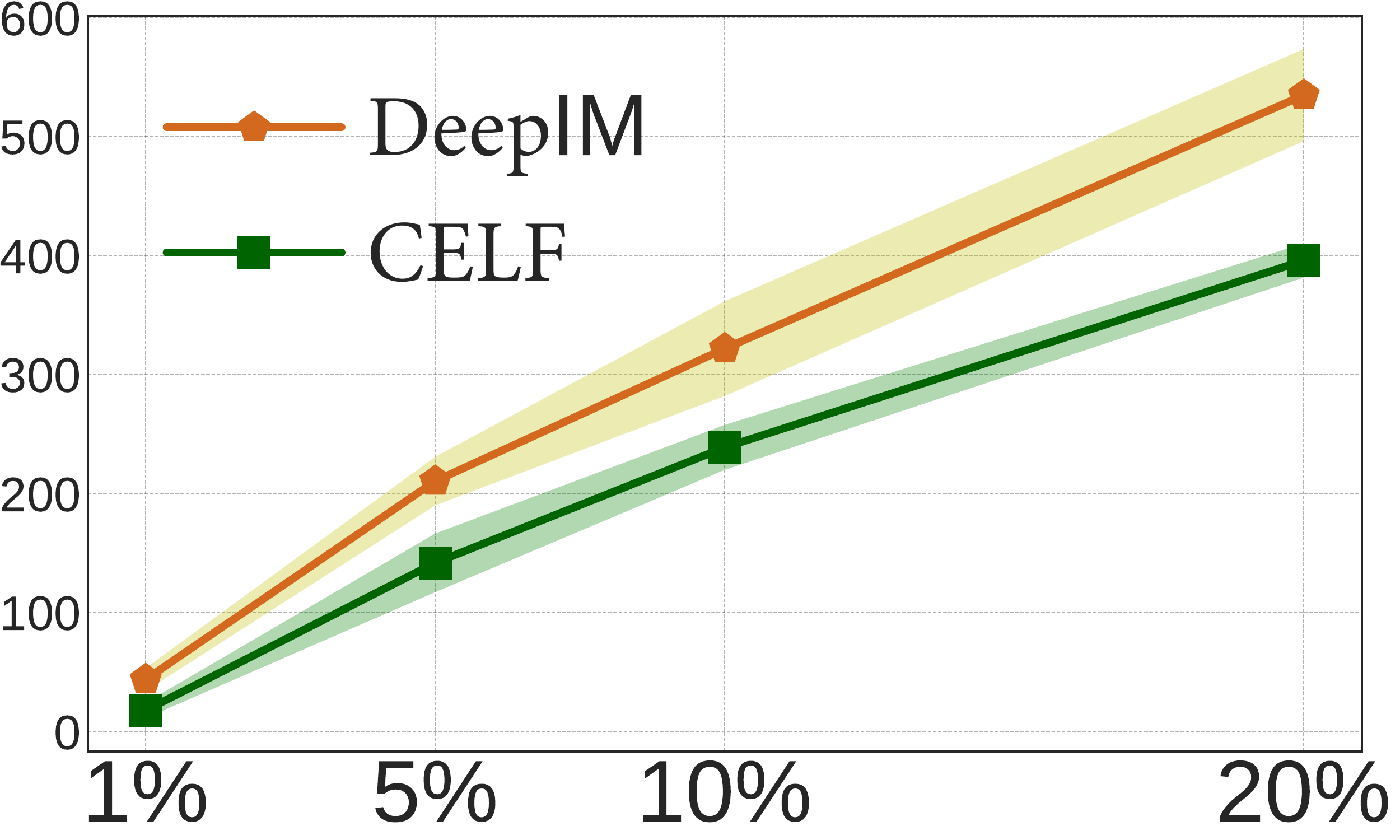}}
		\subfloat[Power Grid-LT]{\label{fig: power_budget_lt}
			\includegraphics[width=0.2\textwidth]{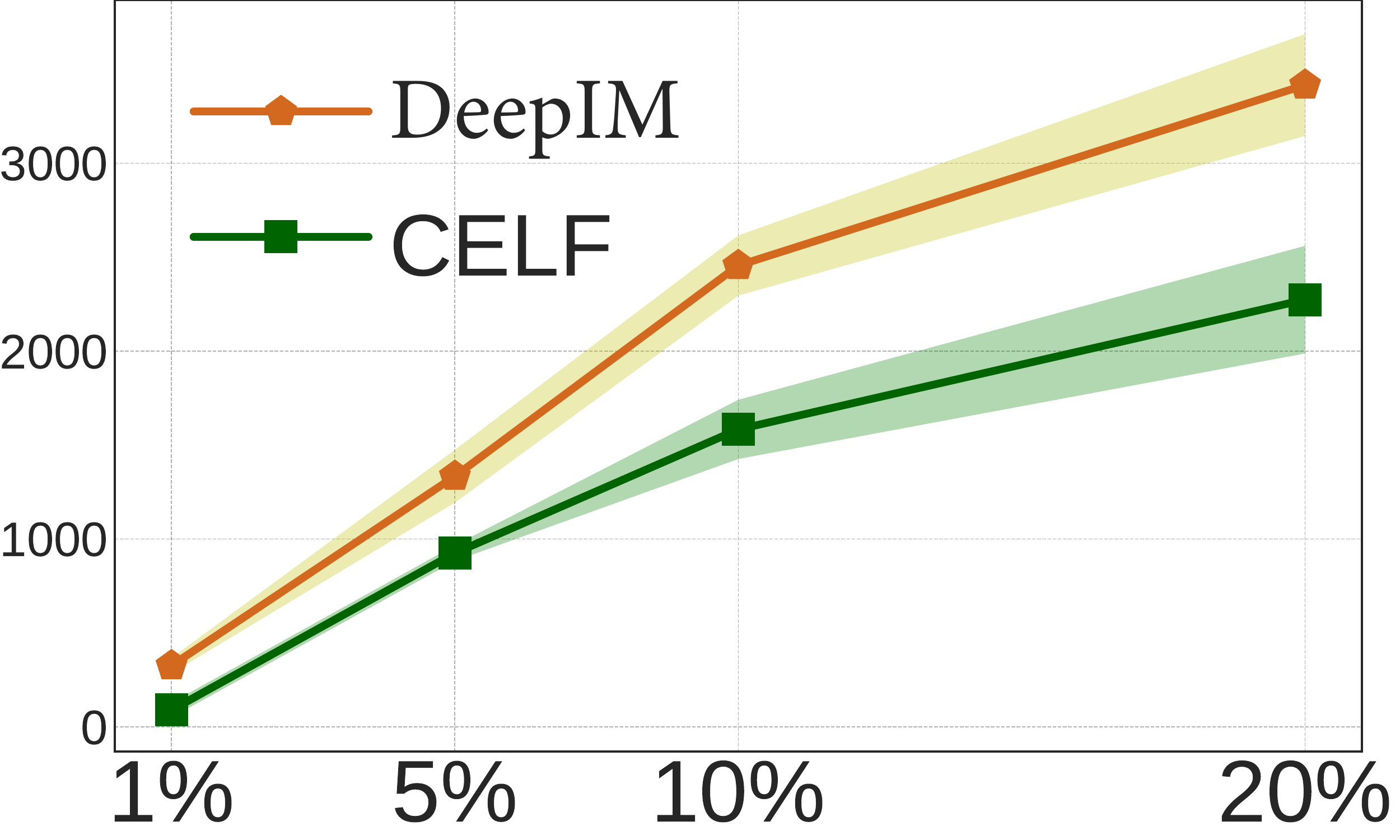}}
		\subfloat[Jazz-LT]{\label{fig: jazz_budget_lt}
			\includegraphics[width=0.2\textwidth]{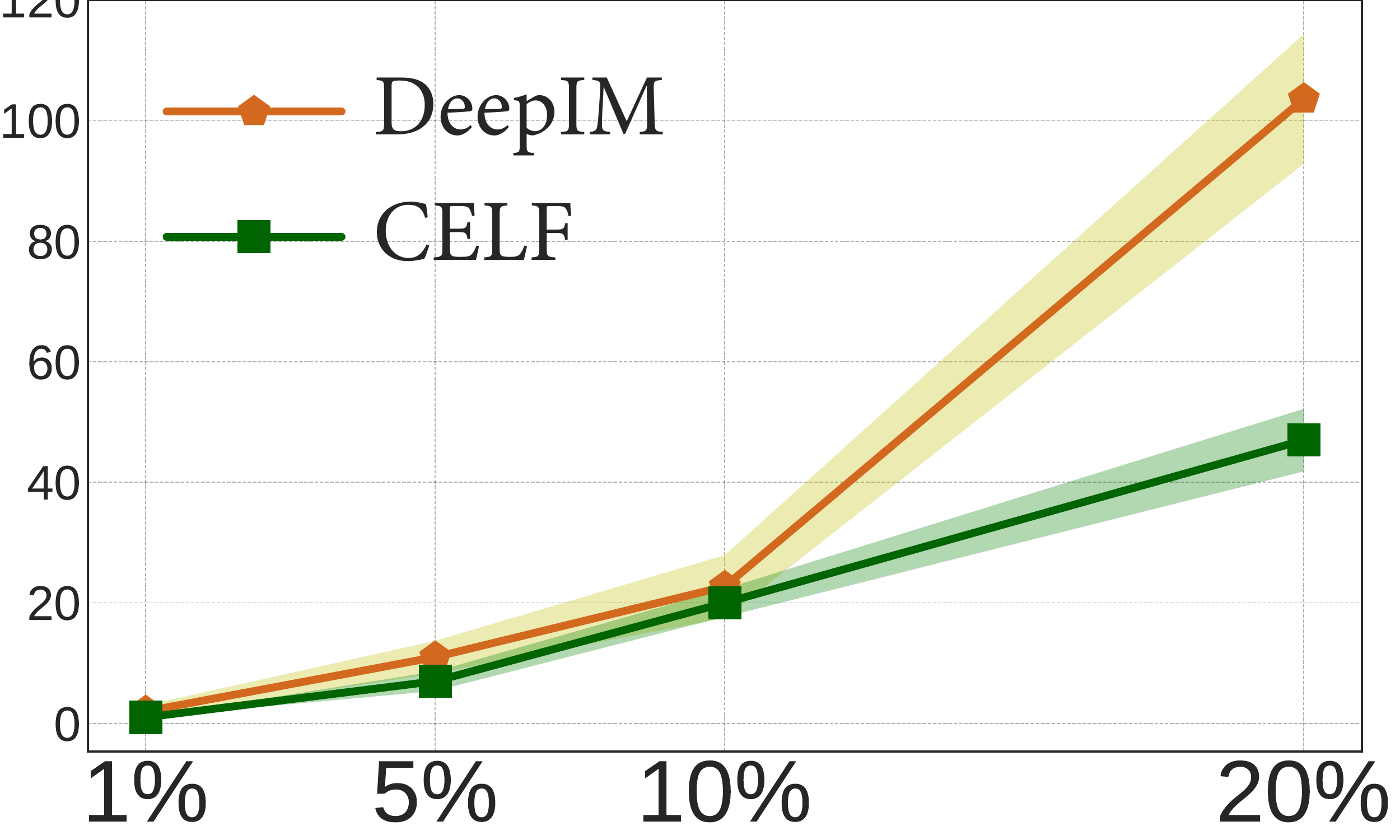}}
		\subfloat[Synthetic-LT]{\label{fig: syn_budget_lt}
			\includegraphics[width=0.2\textwidth]{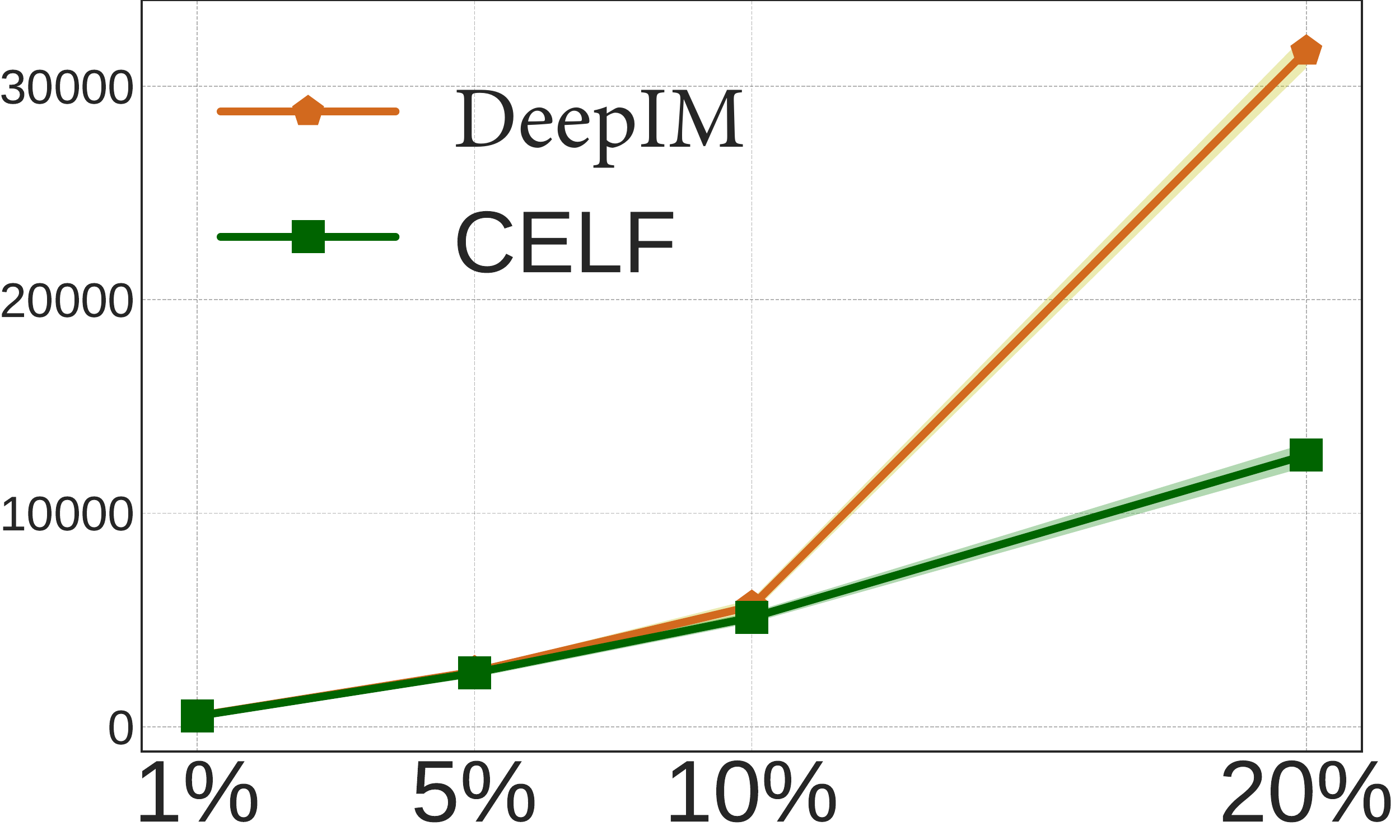}}
			\vspace{-3mm}
		\caption{The influence spread (total infected nodes) in the y-axis under the constraint of the budget with the node size growth (x-axis: 1\%, 5\%, 10\%, and 20\%). Fig. \ref{fig: coraml_budget_ic} - \ref{fig: syn_budget_ic} and Fig. \ref{fig: coraml_budget_lt} - \ref{fig: syn_budget_lt} are evaluated under the IC and LT model, respectively.}
		\vspace{-5mm}
		\label{fig: budget_constraint}
	\end{figure*}
    \subsection{Comparison Method.}
    In addition to comparing our model's performance between the GNN-based diffusion model $M(x,G; \theta)$ (denoted as DeepIM) and the student diffusion model $M_s(z; \lambda)$ (denoted as DeepIM$_s$), we also adopt four sets of comparison methods, all of which are outlined as follows. \textit{Traditional IM}: 1) IMM~\cite{tang2015influence}, 2) OPIM-C~\cite{tang2018online}, and 3) SubSIM~\cite{guo2020influence}. \textit{Learning-based IM}: 1) IMINFECTOR~\cite{panagopoulos2020multi}, 2) PIANO \cite{li2022piano}, and 3) ToupleGDD \cite{chen2022touplegdd}. \textit{Online IM}: OIM~\cite{lei2015online}. \textit{Budget-constraint IM}: CELF~\cite{leskovec2007cost}. In addition, we also compare the performance of the student model DeepIM$_s$ (i.e., coupled with the simplified diffusion model $M_s(\cdot)$).
 
	\subsection{Quantitative Analysis}\label{sec: qua}
	We evaluate the performance of DeepIM in maximizing the influence against other approaches under various IM application schemes. Each model selects 1\%, 5\%, 10\%, and 20\% nodes in each dataset as seed nodes, and we allow each diffusion model to simulate until the diffusion process stops and record the average influence spread of $100$ rounds. We report the percentage of final infected nodes (i.e., the number of infected nodes/the total number of nodes).

	\begin{figure*}[!t]
 \vspace{-3mm}
 \centering
		\subfloat[DeepIM-20\%]{\label{fig: visual_DeepIM_20}
			\hspace{-3mm}\includegraphics[width=0.2\textwidth, trim = 0cm 5cm 0cm 4cm, clip]{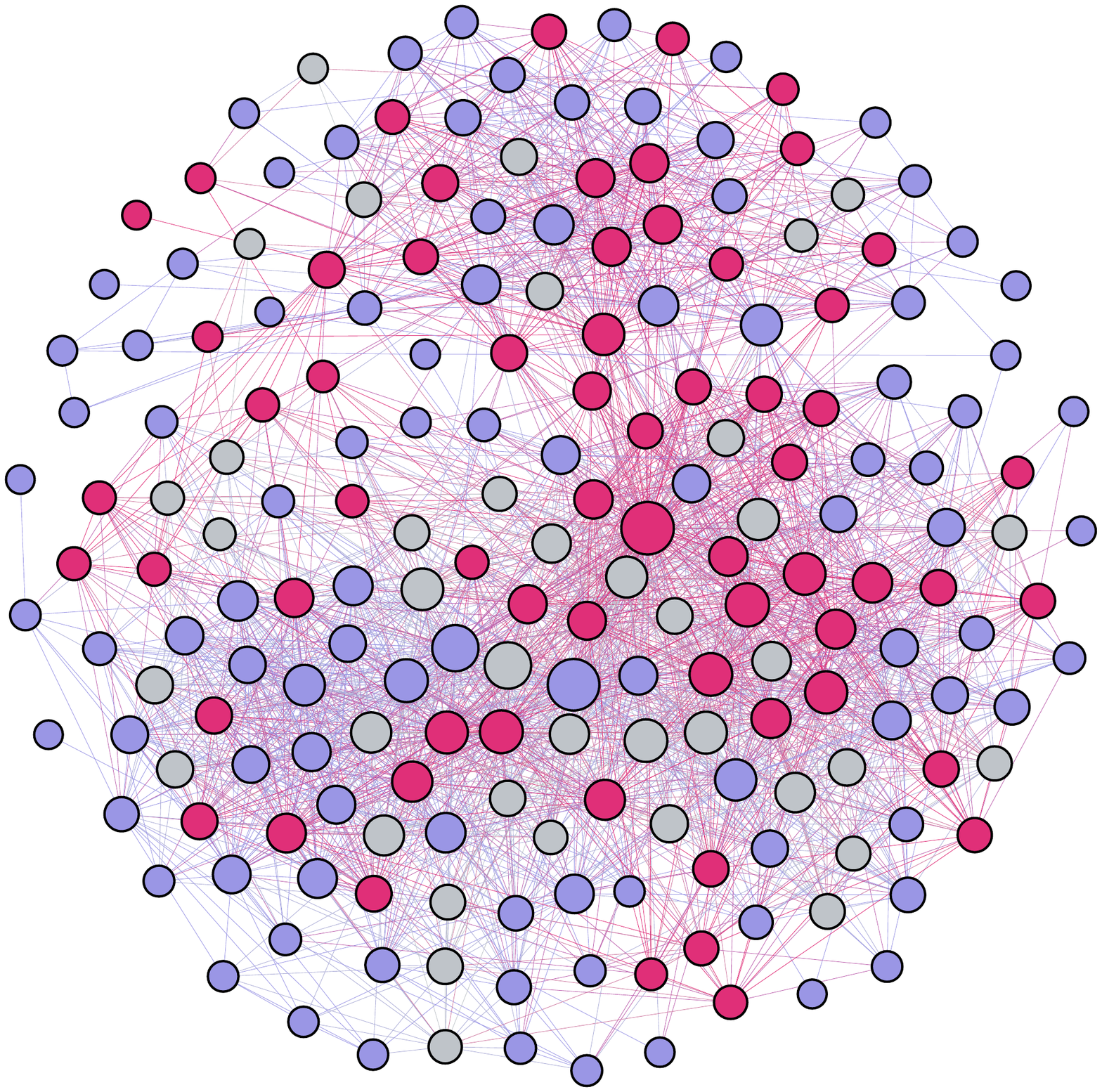}}
		\subfloat[OIM-20\%]{\label{fig: visual_greedy_20}
			\hspace{-3mm}\includegraphics[width=0.2\textwidth, trim = 0cm 5cm 0cm 4cm, clip]{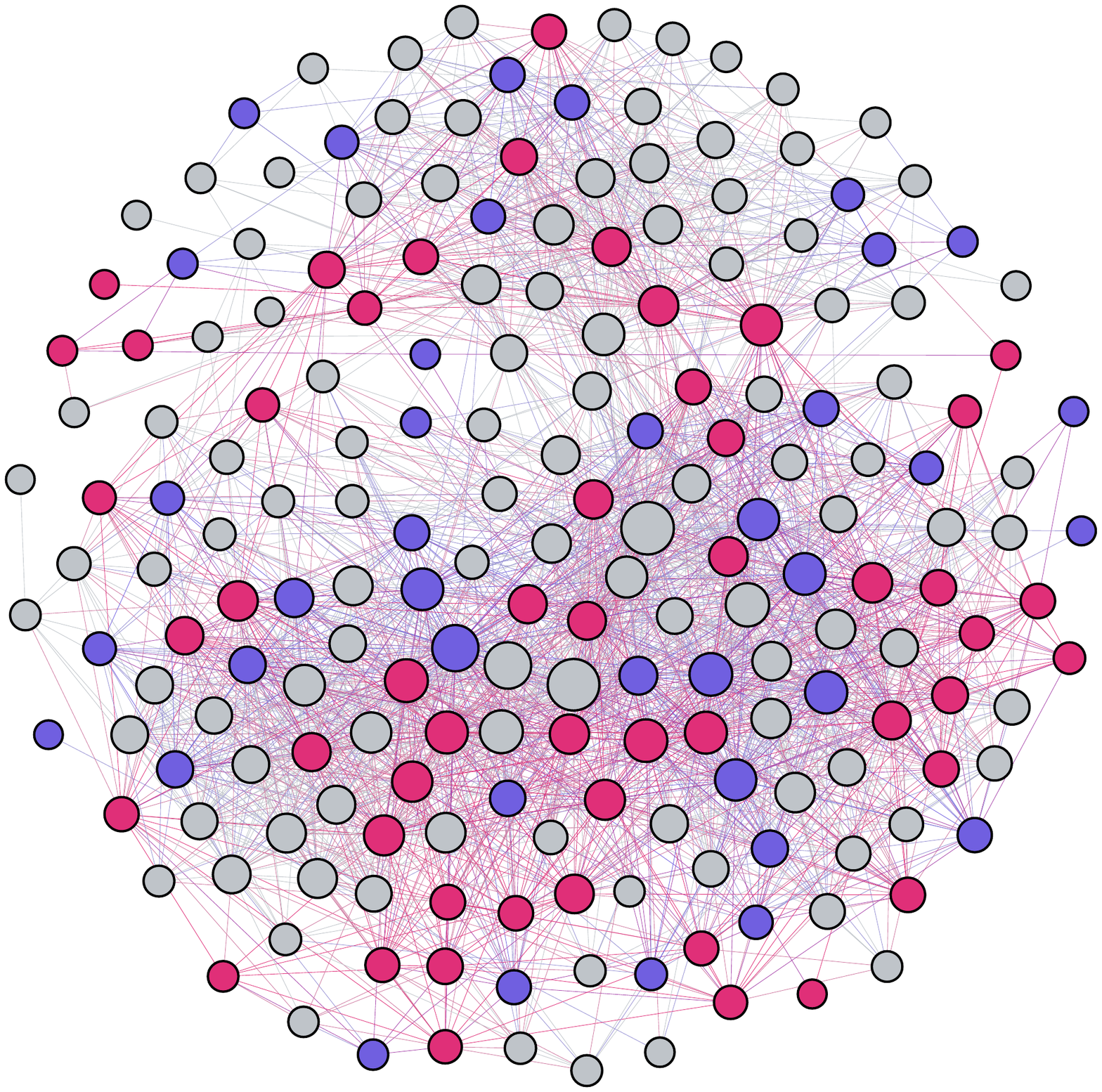}}
		\subfloat[OPIM-20\%]{\label{fig: visual_imm_20}
			\hspace{-3mm}\includegraphics[width=0.2\textwidth, trim = 0cm 5cm 0cm 4cm, clip]{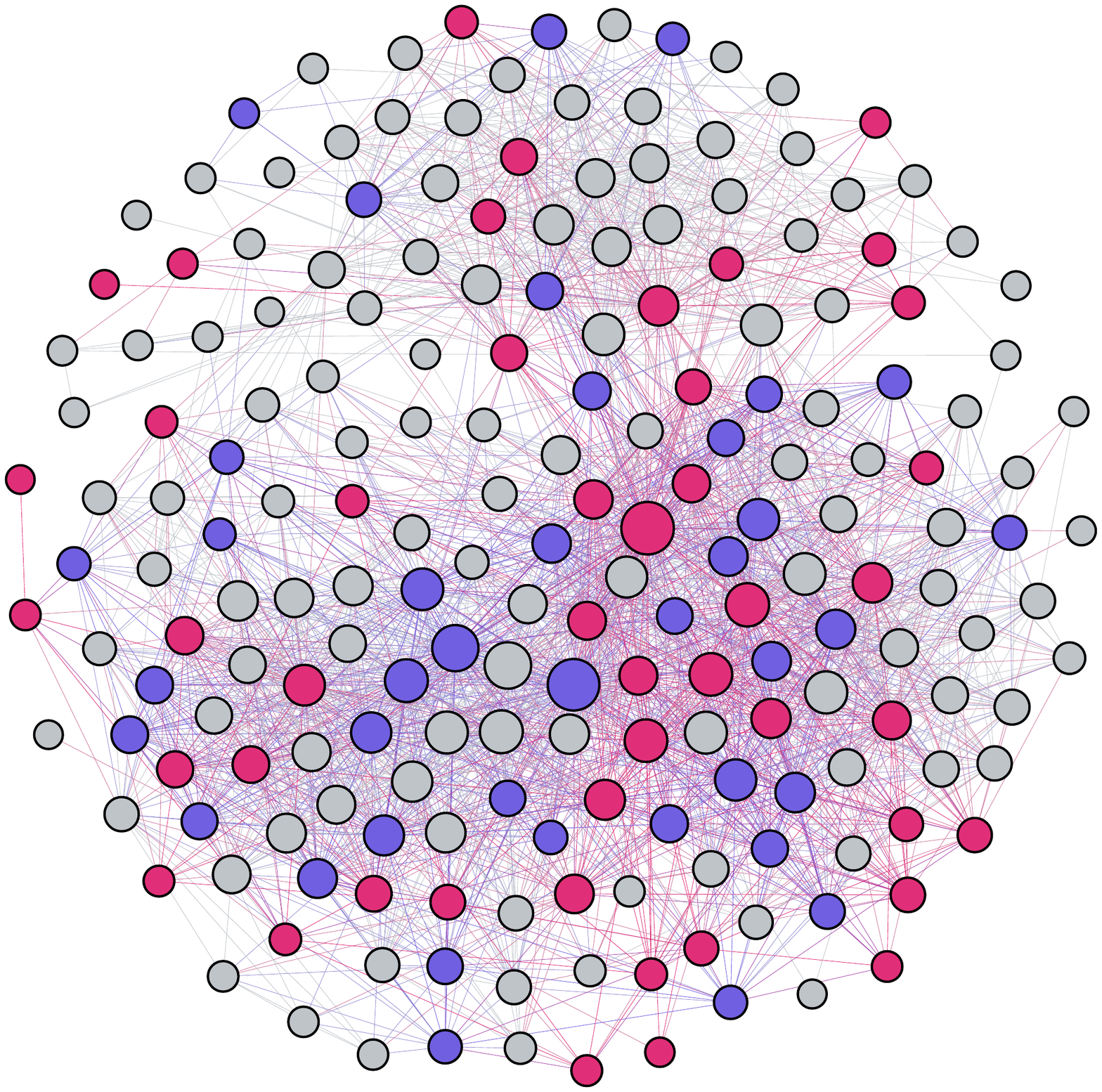}}
		\subfloat[SubSIM-20\%]{\label{fig: visual_opim_20}
			\hspace{-3mm}\includegraphics[width=0.2\textwidth, trim = 0cm 5cm 0cm 4cm, clip]{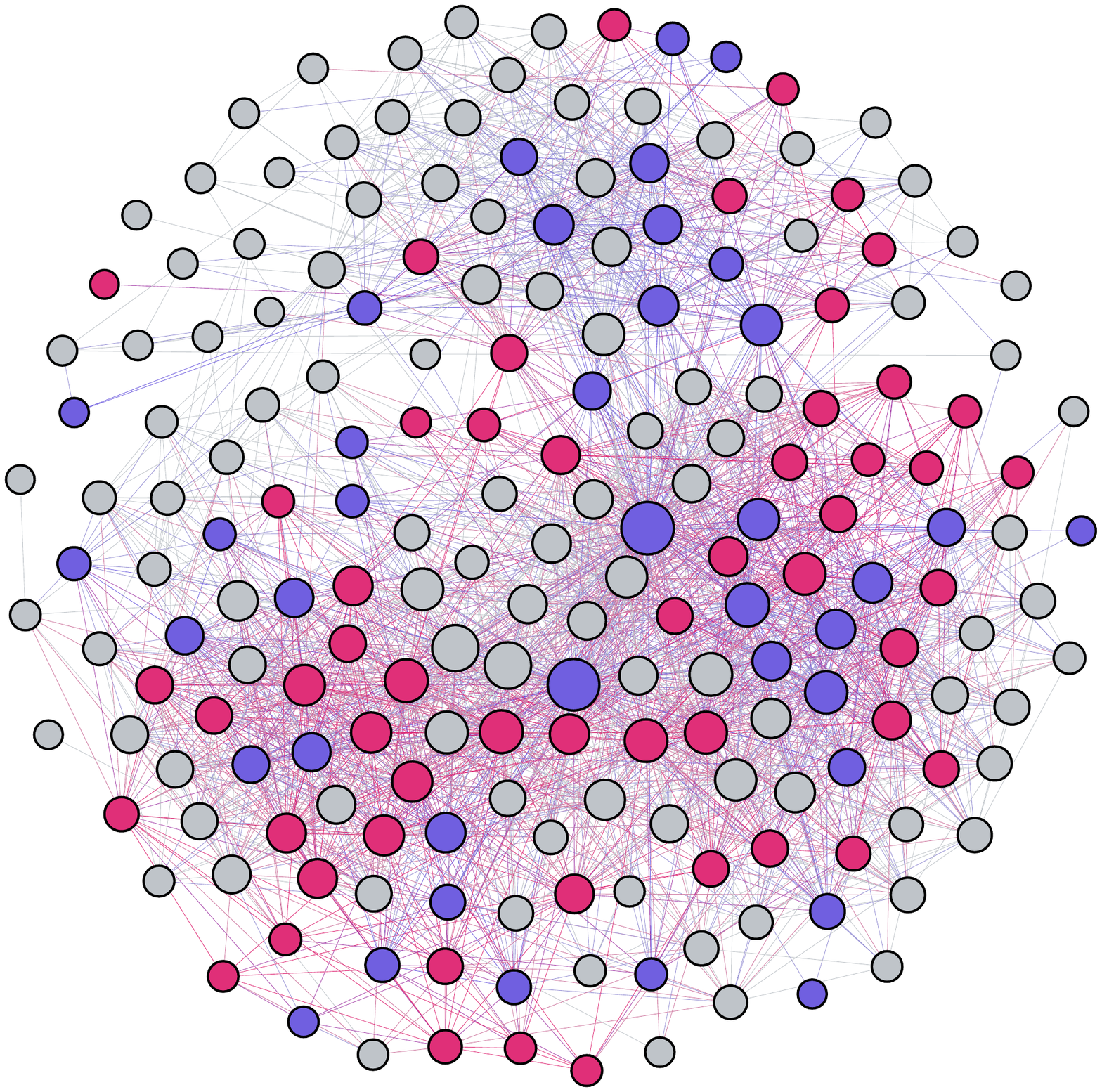}}
		\subfloat[ToupleGDD-20\%]{\label{fig: visual_sub_20}
			\hspace{-3mm}\includegraphics[width=0.2\textwidth, trim = 0cm 5cm 0cm 4cm, clip]{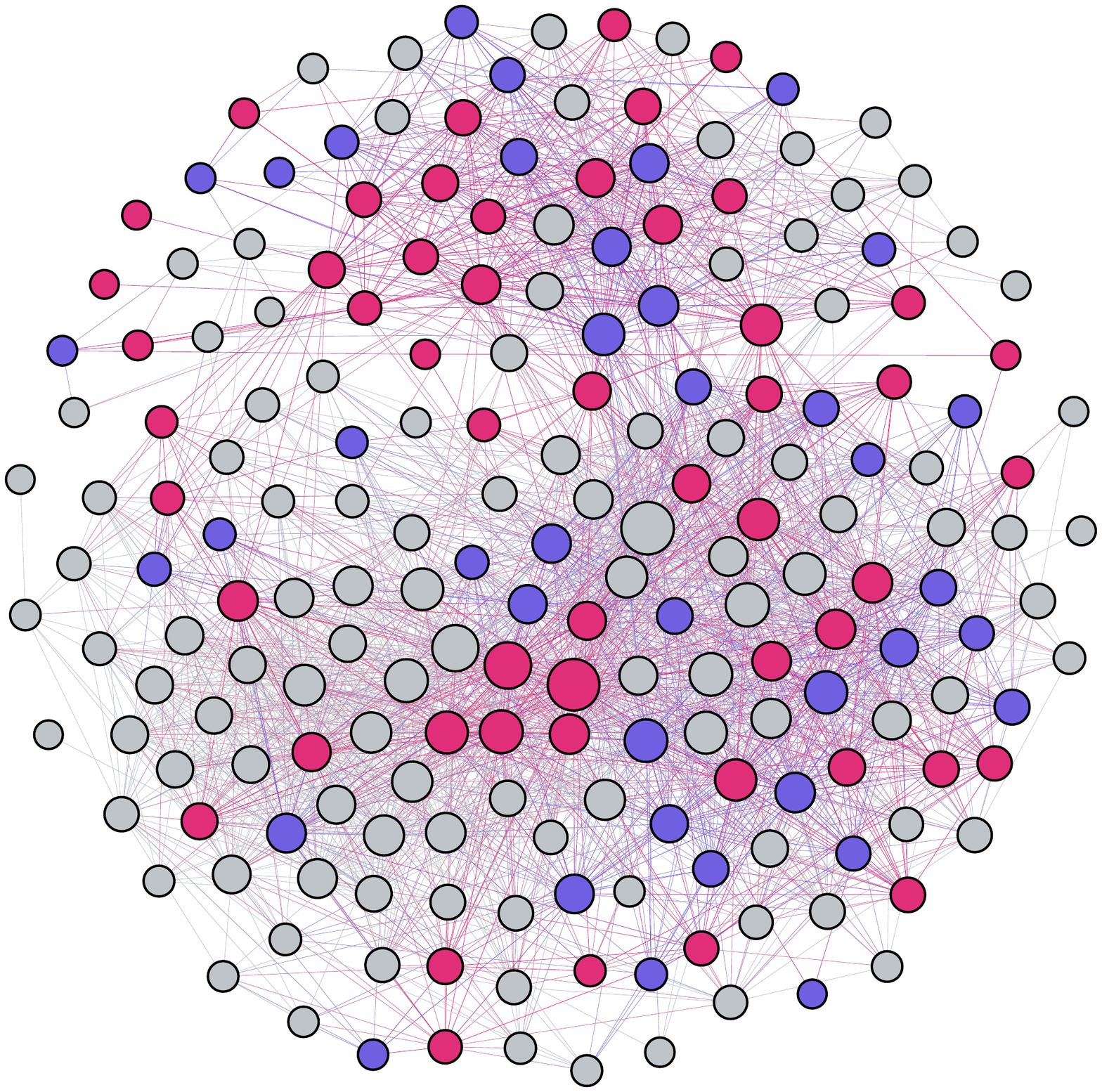}}
		\vspace{-3mm}
		\caption{The visualization of influence spread in Jazz dataset: The size of nodes is determined by the node degree, and the color on nodes determines the infection status: blue means the node is in seed set, red means the node is infected, and grey means the node is not infected.}
		\vspace{-5mm}
		\label{fig: visualization}
	\end{figure*}
	\noindent\textbf{IM under IC Model.} We first examine the effectiveness of DeepIM against other baseline methods under the IC diffusion pattern. As shown in Table \ref{tab: evaluation_ic}, DeepIM can achieve an overall better performance than other methods across all datasets. Compared to traditional methods, IMM, OPIM, and SubSIM are three state-of-the-art methods based on reserve-set sampling and various approximation techniques, which generate similar results across all datasets; however, they rely on different heuristics to guide the node selection for efficiency improvement and fail to decode the underlying distribution of seed sets. OIM achieves better performance than traditional methods in most datasets because it can automatically update the edge weight iteratively. However, the disadvantage of OIM is also obvious: it is tailored for the specific IC diffusion model, which is less applicable in real-world scenarios. Lastly, the learning-based IM methods (IMINFECTOR, PIANO, and ToupleGDD) achieve competitive and generally better performance than traditional ones due to their larger model size and better generalization ability. However, learning-based methods that leverage reinforcement learning experience scalability issues and they cannot be applied in billion-scale networks (e.g., Digg and Weibo), which makes them hard to be applied in real-world scenarios. Compared to learning-based methods, DeepIM proposes a more robust way of learning the end-to-end diffusion model and searching the high-influential node set in latent space directly, which can better capture the underlying diffusion dynamics and resolve the scalability issues. In addition, DeepIM$_s$ incorporates a lightweight end-to-end learning-based diffusion model that could retain both efficacy and efficiency than other learning-based methods.
 
	\noindent\textbf{IM under LT Model.} We then evaluate the final influence spread with respect to the initial seed set size by assuming LT as the diffusion model. As shown in Table \ref{tab: evaluation_lt}, DeepIM can generate more superior seed sets to infect the most number of nodes and excel other methods by an evident margin across all datasets. Notably, DeepIM demonstrates its superiority in the Synthetic dataset that can effectively spread the influence to the whole network when choosing $20\%$ of the node as the initial seed set, yet other methods can infect at most $70\%$ nodes of the network. Specifically, DeepIM and DeepIM$_s$ excel other methods by on average $200\%$ in the Jazz dataset and $30\%$ in the Synthetic dataset. The reason is largely due to the lack of generalization capability of other methods under various diffusion models.

	\textbf{IM with Budget Constraint.} 
	We then compare the quality of the seed sets generated by DeepIM and CELF under the IC and LT model with the budget constraint, and such a budget is explicitly defined as the node degree in this paper. As can be seen from Fig. \ref{fig: budget_constraint}, our proposed method generally performs better than CELF across all networks of different sizes, and the margins are more evident under the LT model (Fig. \ref{fig: coraml_budget_lt} - \ref{fig: syn_budget_lt}). In addition, compared to CELF, the growths of influence spread in DeepIM have fewer fluctuations across all datasets, which also demonstrates the stability of DeepIM because of its capability of identifying the latent distribution seed sets while considering the budget constraint.

 \begin{table}[t]
    \centering
    \resizebox{0.9\columnwidth}{!}{%
    \begin{tabular}{@{}l|cccc|c@{}}
\toprule
           & \multicolumn{1}{l}{10,000} & 20,000  & 30,000  & 50,000  & \multicolumn{1}{l}{50,000 (Training)}                                       \\ \midrule
IMINFECTOR & 3.478s                     & 7.842s  & 12.376s & 16.492s & 4753.67s                                                                    \\
PIANO      & 5.948s                     & 10.532s & 16.575s & 28.437s & 14732.63s                                                                   \\
ToupleGDD  & 10.476s                    & 19.583  & 32.792s & 58.985s & --                                                                          \\ \midrule
DeepIM$_s$ & 0.312s                     & 0.616s  & 0.847s  & 1.275s  & 503.12s                                                                     \\
DeepIM     & 1.402s                     & 2.798s  & 5.124s  & 12.882s & 1244.56s \\ \bottomrule
\end{tabular}%
    }
    \vspace{-3mm}
    \caption{The average inference runtime (in seconds) with regard to the increase of node size and the average training time. We select $10\%$ of nodes as the seeds.}
    \label{tab: time}
    \vspace{-7mm}
    \end{table}
    
    \subsection{Scalability Analysis}
	We record the runtime of the seed set inference with regard to the increase in node size against other learning-based IM solutions. As can be clearly seen in Table \ref{tab: time}, DeepIM demonstrates near-linear growth of runtime as the graph size increases. In addition, it achieves a generally shorter inference time (on average $20\%$ faster inference time than the second-fast IMINFECTOR) compared to other learning-based methods. In addition, our DeepIM$_s$ coupled with a lightweight end-to-end diffusion model can greatly reduce the computational cost of estimating the expected influence spread, and achieves even $90\%$ improvement in the inference time on average than our DeepIM model.

	\subsection{Case Study: Graph Diffusion Visualization}
    Finally, we conduct a case study to demonstrate the distribution of selected $20\%$ seed nodes as well as the final infection status of all nodes in Fig. \ref{fig: visualization}, where blue nodes indicate the initial seed node, red nodes indicate the infected node during the influence spread, and grey nodes represent uninfected nodes. Due to the ease of representation, we only visualize the result of the Jazz dataset because of its overall smaller graph size. Overall, DeepIM demonstrates better performance in terms of spreading influence. Due to the space limit, we compare the influence spread result between different initial seed set sizes, namely $10\%$ and $20\%$, in the appendix and provide more discussions there.
    	
	\section{Conclusion}\label{sec:con}
	In this paper, we propose a novel framework to tackle the IM problem in a more robust and generalized way than existing learning-based IM methods. Particularly, to characterize the complex nature of the seed set, we propose to character the probability of the seed set and directly search for a more optimal seed set in continuous space. Furthermore, to solve the challenge of modeling the underlying diffusion pattern, we offer two different learning-based diffusion models to characterize the diversified diffusion dynamics with efficiency and efficacy guarantee. Finally, we propose a novel objective function that can be coupled with multiple constraints for seed node set inference, which can adapt to different IM application schemes. Extensive experiments and case studies on both synthetic and real-world datasets demonstrate the advantages of DeepIM over existing state-of-the-art methods to maximize the influence spread.

\clearpage

\bibliography{main}
\bibliographystyle{icml2023}

\appendix
\section{Proofs}\label{sec: proofs}
The proof of Theorem \ref{theorem: submodular} is demonstrated as follows.
\begin{proof}
$g_u(x, G;\theta)=\mathcal{A}^1\circ (C^{1} \circ \mathcal{A}^2\circ C^{2}\cdots\circ \mathcal{A}^K\circ C^K)$ via iterating Eq. \eqref{eq:GNN} recursively. Because $\mathcal{A}^k$ and $C^k$ are  non-decreasing, so is $\mathcal{A}^1\circ C^{1}\cdots\circ \mathcal{A}^K\circ C^K$, which is $g_u$. Therefore, $M$ is infection monotonic. Because $g_u$ and $g_r$ are non-decreasing, $M$ is also non-decreasing and hence score monotonic.
	\end{proof}

\begin{table*}[t]
    \centering
    \resizebox{\textwidth}{!}{%
    \begin{tabular}{@{}c|cccc|cccc|cccc|cccc|cccc|cccc|cccc@{}}
    \toprule
        & \multicolumn{4}{c|}{Cora-ML}         & \multicolumn{4}{c|}{Network Science}      & \multicolumn{4}{c|}{Power Grid}   &  \multicolumn{4}{c|}{Jazz} &
        \multicolumn{4}{c|}{Synthetic}
        &
        \multicolumn{4}{c|}{Digg}
        &
        \multicolumn{4}{c|}{Weibo}\\ \midrule
    Methods & 1\%     & 5\%     & 10\%     & 20\%    & 1\%     & 5\%     & 10\%     & 20\%    & 1\%     & 5\%     & 10\%     & 20\%    & 1\%     & 5\%     & 10\%     & 20\%    & 1\%     & 5\%     & 10\%     & 20\%
    & 1\%     & 5\%     & 10\%     & 20\%
    & 1\%     & 5\%     & 10\%     & 20\%\\ \midrule
    Greedy         & 1.6           & 8.3          & 14.8          & 26.1          & 1.1          & 5.4           & 11.6          & 20.8          & 1.1          & 5.0          & 10.2          & 21.3          & 17.4          & 33.7           & 49.6           & 64.2          & 2.5          & 12.1          & 19.5           & 35.5 
    & 1.9          & 8.6          & 15.6           & 31.2 
    & 1.5          & 7.2          & 13.9           & 28.7 \\
    IMINFECTOR         & 2.1          & 9.4          & 16.1          & 27.9          & 1.7           & 5.8          & 12.4          & 22.3          & 1.3          & 5.5           & 12.4          & 23.1          & 8.8          & 35.4          & 54.8          & 66.2           & 2.5          & 12.4          & 20.5          & 36.2
    & 2.3          & 9.1          & 16.4         & 32.4
    & 2.5         & 8.5          & 15.5          & 29.6\\
    IMM            & 2.0           & 9.5          & 15.4          & 27.6          & 1.3          & 5.6           & 12.2          & 22.1          & 1.1          & 5.6          & 11.0          & 22.9          & 7.6          & 37.8           & 55.6          & 67.1          & 2.7          & 12.6          & 20.9          & 37.3          & 2.5          & 9.4          & 16.3           & 32.6 
    & 2.3          & 8.1          & 15.7           & 29.4 \\
    OPIM           & 2.3           & 9.3          & 16.2          & 27.2          & 1.4          & 5.9           & 13.0          & 22.1          & 1.2          & 5.9          & 11.2          & 22.4          & 5.7          & 44.7           & 58.6          & 68.3          & 2.8          & 12.5          & 20.2          & 36.1          & 2.3          & 9.3          & 16.5           & 32.1 
    & 2.3          & 8.5          & 15.3           & 29.7 \\
    SubSIM         & 2.3           & 9.2          & 16.9          & 28.8          & 1.5          & 5.6           & 12.2          & 23.3          & 1.2          & 5.6          & 11.4          & 21.9          & 2.9           & 30.1           & 53.8          & 67.0          & 2.5          & 12.6          & 20.2          & 36.5          & 2.5          & 9.5          & 16.1           & 32.3 
    & 2.3          & 8.3          & 15.6           & 29.4 \\
    \textbf{DeepIM} & \textbf{7.1} & \textbf{16.1} & \textbf{21.9} & \textbf{30.8} & \textbf{2.7} & \textbf{8.7} & \textbf{15.1} & \textbf{25.1} & \textbf{1.9} & \textbf{7.6} & \textbf{13.3} & \textbf{23.8} & \textbf{27.1} & \textbf{57.1} & \textbf{68.1} & \textbf{74.1} & \textbf{3.2} & \textbf{14.4} & \textbf{24.5} & \textbf{39.1}  & \textbf{5.6}          & \textbf{11.4}          & \textbf{18.8}           & \textbf{36.3} 
    & \textbf{6.5}          & \textbf{13.1}          & \textbf{17.1}           & \textbf{32.3} \\
    \bottomrule
    \end{tabular}
    }
    \vspace{-3mm}
    \caption{Performance over comparison methods under SIS diffusion pattern. (Best is highlighted with bold.)}
    \vspace{-3mm}
    \label{tab: evaluation_sis}
    \end{table*}

The proof of Corollary \ref{theorem: gcn_gat} is demonstrated as follows.
    \begin{proof}
    For the GAT model, $a_{i,j}^k=\mathcal{A}^k(h_{i}^k,h_{j}^k,\theta^k)=\theta^k_1(\theta^k_2h^{k-1}_i||\theta^k_2 h^{k-1}_j)$ and $h_i^k=C^k(a^k,\theta^k)=\max(\sum\nolimits_{j\in N(i)} softmax(LeakyReLU(a^k_{i,j}))\theta^k_1 h^{k-1}_j,0)$ where $\theta^k=[\theta^k_1;\theta^k_2]$, and $||$ denotes the concatenation of two vectors. $\mathcal{A}^k$ are non-decreasing and non-negative because $\theta^k\geq 0$ (i.e. $\theta^k_1\geq 0$ and $\theta^k_2\geq 0$). In other words, $a^k_{i,j}$ is non-negative, and $h_i^k$, $\theta^k_1$ and the softmax operator are non-negative. Therefore, the LeakyReLU operator and $\max(\bullet,0)$ can be removed from $C^k$. That is, $h_i^k=C^k(a^k,\theta^k)=\sum\nolimits_{j\in N(i)} softmax(a^k_{i,j})\theta^k_1 h^{k-1}_j$. Because the softmax operator is non-decreasing, and $\theta^k_1$ is non-negative. $C^k$ is non-decreasing. Hence, The GAT model satisfies the conditions in Theorem \ref{theorem: submodular} and thus $M$ is score and infection monotonic.
    \end{proof}

    The proof of Corollary \ref{theorem: consistency} is illustrated as follows. 
    \begin{proof}
    According to Theorem \ref{theorem: submodular}, the GNN-based $M(x, G; \theta)$ is monotonic. Then for any two $x^{(i)} > x^{(j)}$, we have $M(x^{(i)}, G; \theta) > M(x^{(j)}, G; \theta)$. If the reconstruction error is minimized during the training of $f_{\psi}(\cdot)$, we also have $f_{\psi}(z^{(i)}) > f_{\psi}(z^{(j)})$. Hence, $M(f_{\psi}(z^{(i)}), G; \theta) > M(f_{\psi}(z^{(j)}), G; \theta)$ also holds.
    \end{proof}

    The derivation of Equation \ref{eq: infer_2} is shown as follows.
    \begin{align*}
        \mathcal{L}_{\text{pred}}
        &= \min_z \mathbb{E}\big[- \log p_{\theta}(y|x, G) - \log p_{\psi}(x|z)\big],\\ 
        &\hspace{5mm}\text{s.t.} \sum\nolimits^{|V|}_{i=0} \mathcal{F}(v_i, G)\cdot x_i \le k,\\
    \end{align*}
    Since we assume the optimal $\Tilde y = |V|$ and the predicted $y \in [0, |V|]$, the optimization target is to maximize the $y$ until it reaches the fully infected status. Therefore, the first term in  Equation \eqref{eq: infer_2} is written as the Mean Squared Loss (MSE): $\norm{\Tilde y-M(x, G; \theta)}_2^2$. For the second term in Equation \eqref{eq: infer_2}, the value range of $x$ after the autoencoder is $[0, 1]$, indicating the probability of each node being selected to the seed set. $x\in [0, 1]$ fits the binomial distribution so that minimizing the negative log-likelihood is equivalent to minimizing the probability mass function. Therefore, the second term of the above function can be written to $- \log \big[\prod_i^{|V|}f_{\psi}(z_i)^{x_i}(1-f_{\psi}(z_i)^{1-x_i}\big]$. Adding both terms gives us the final expression as shown in Equation \eqref{eq: infer_2}:
    \begin{align*}
        \mathcal{L}_{\text{pred}} &= \max_z \mathbb{E}\big[ p_{\theta}(y|x, G)\cdot p_{\psi}(x|z)\big], \\
        &\hspace{5mm}\text{s.t.} \sum\nolimits^{|V|}_{i=0}  \mathcal{F}(v_i, G)\le k.
    \end{align*}

   \begin{figure*}[!t]
 \vspace{-3mm}
 \centering
        \subfloat[DeepIM-10\%]{\label{fig: visual_DeepIM_10}            \hspace{-3mm}\includegraphics[width=0.2\textwidth, trim = 0cm 5cm 0cm 5cm, clip]{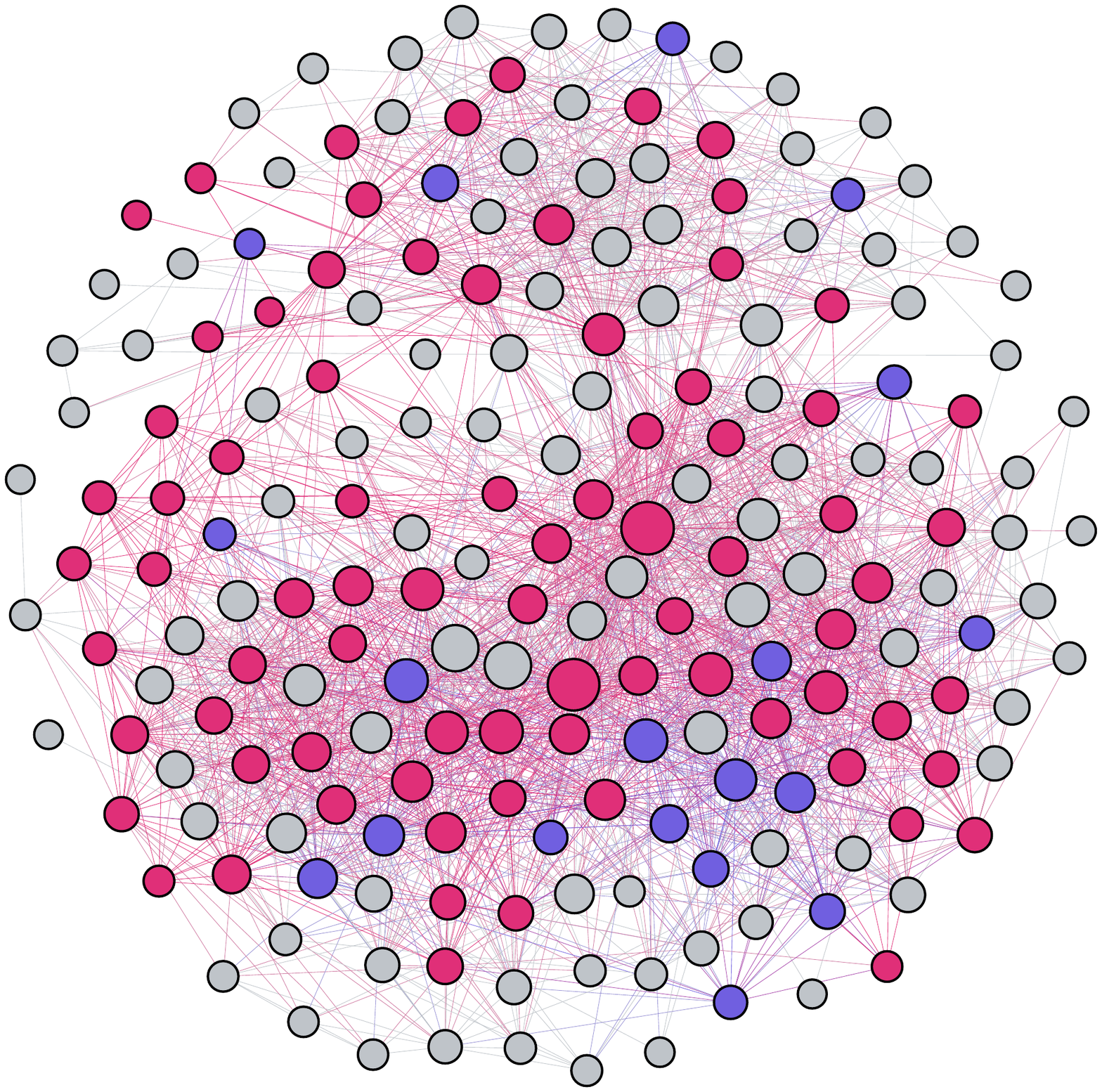}}
        \subfloat[OIM-10\%]{\label{fig: visual_greedy_10}
            \hspace{-3mm}\includegraphics[width=0.2\textwidth, trim = 0cm 5cm 0cm 5cm, clip]{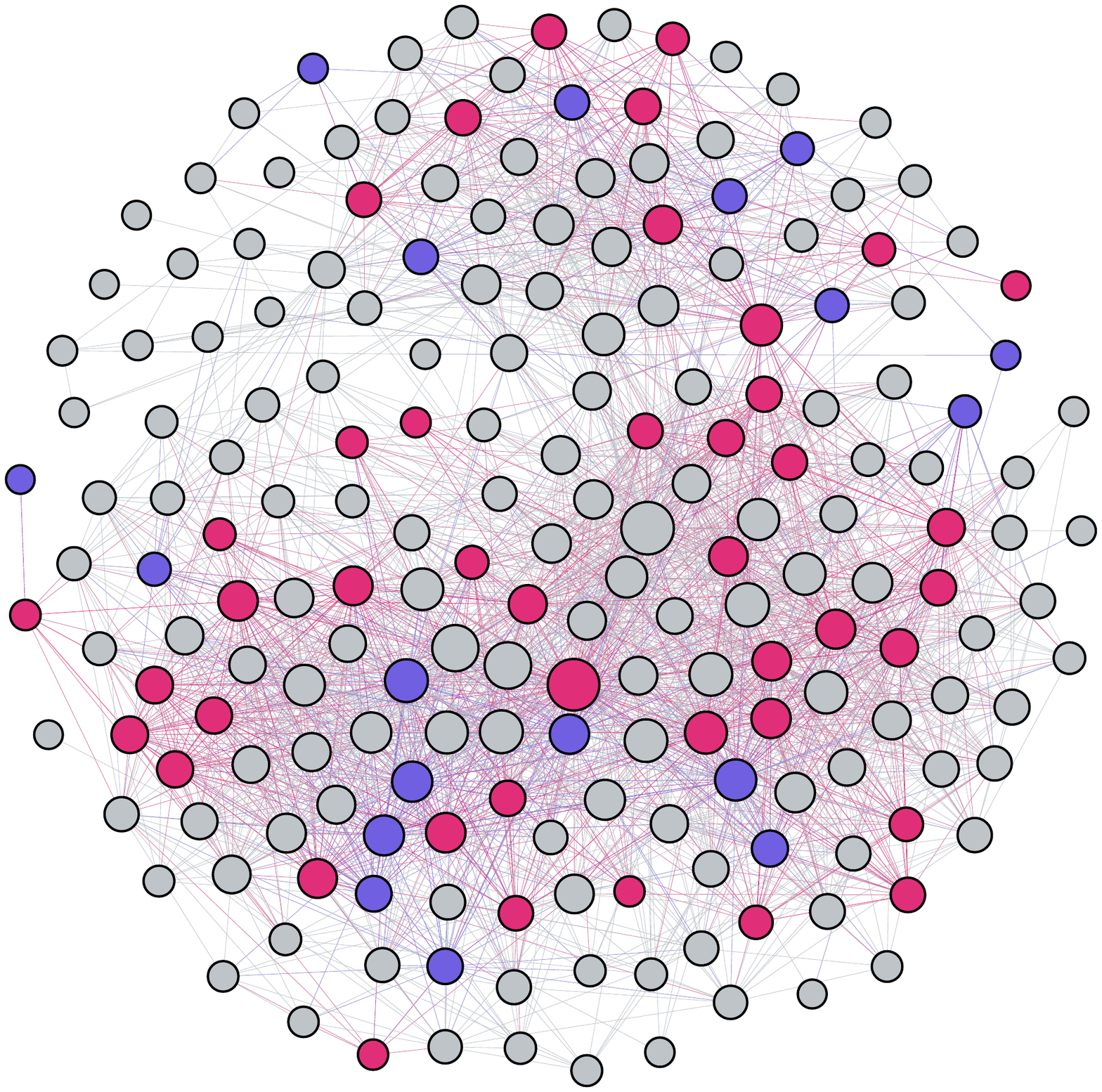}}
        \subfloat[OPIM-10\%]{\label{fig: visual_imm_10}
            \hspace{-3mm}\includegraphics[width=0.2\textwidth, trim = 0cm 5cm 0cm 5cm, clip]{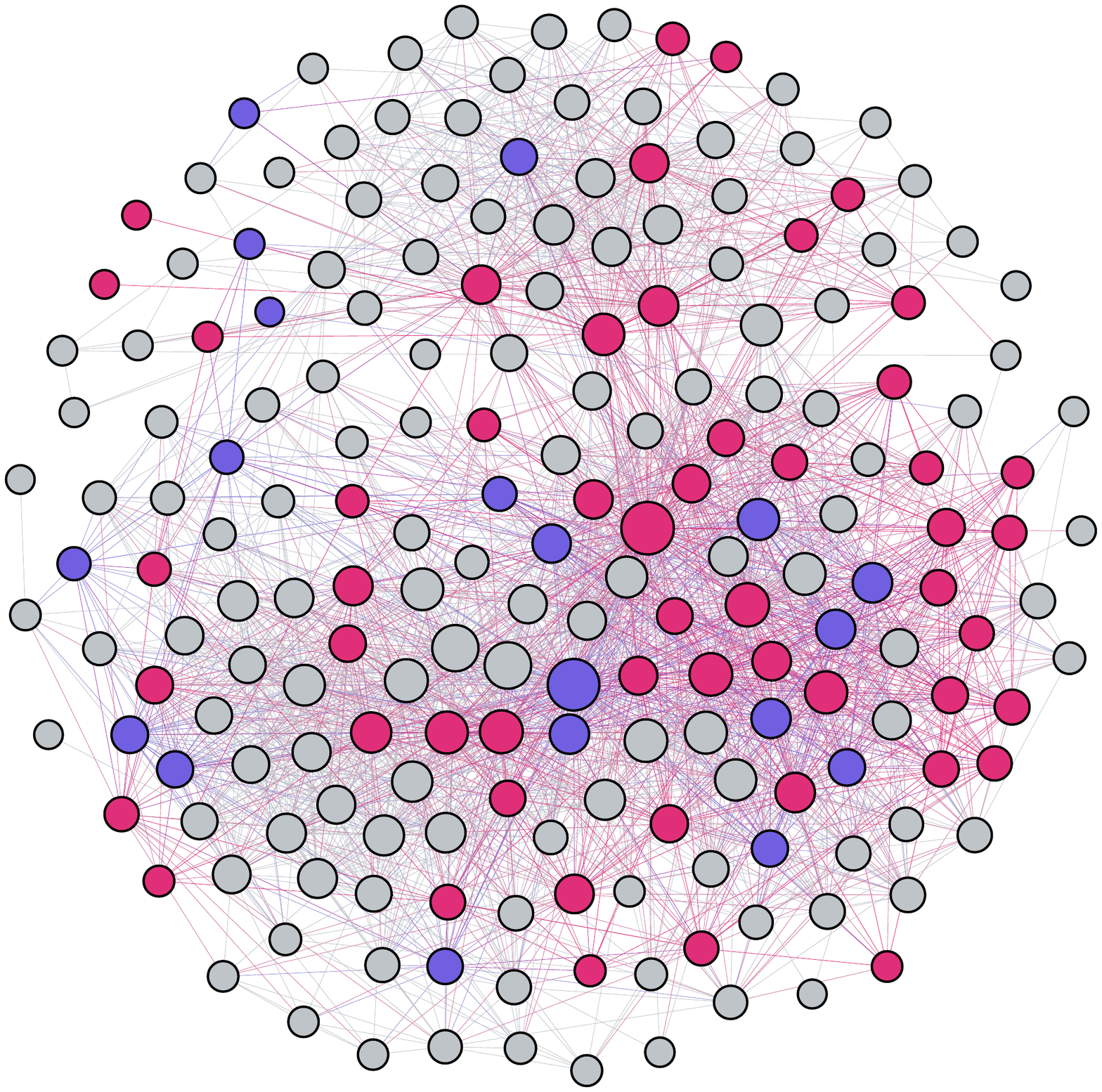}}
        \subfloat[SubSIM-10\%]{\label{fig: visual_opim_10}
            \hspace{-3mm}\includegraphics[width=0.2\textwidth, trim = 0cm 5cm 0cm 5cm, clip]{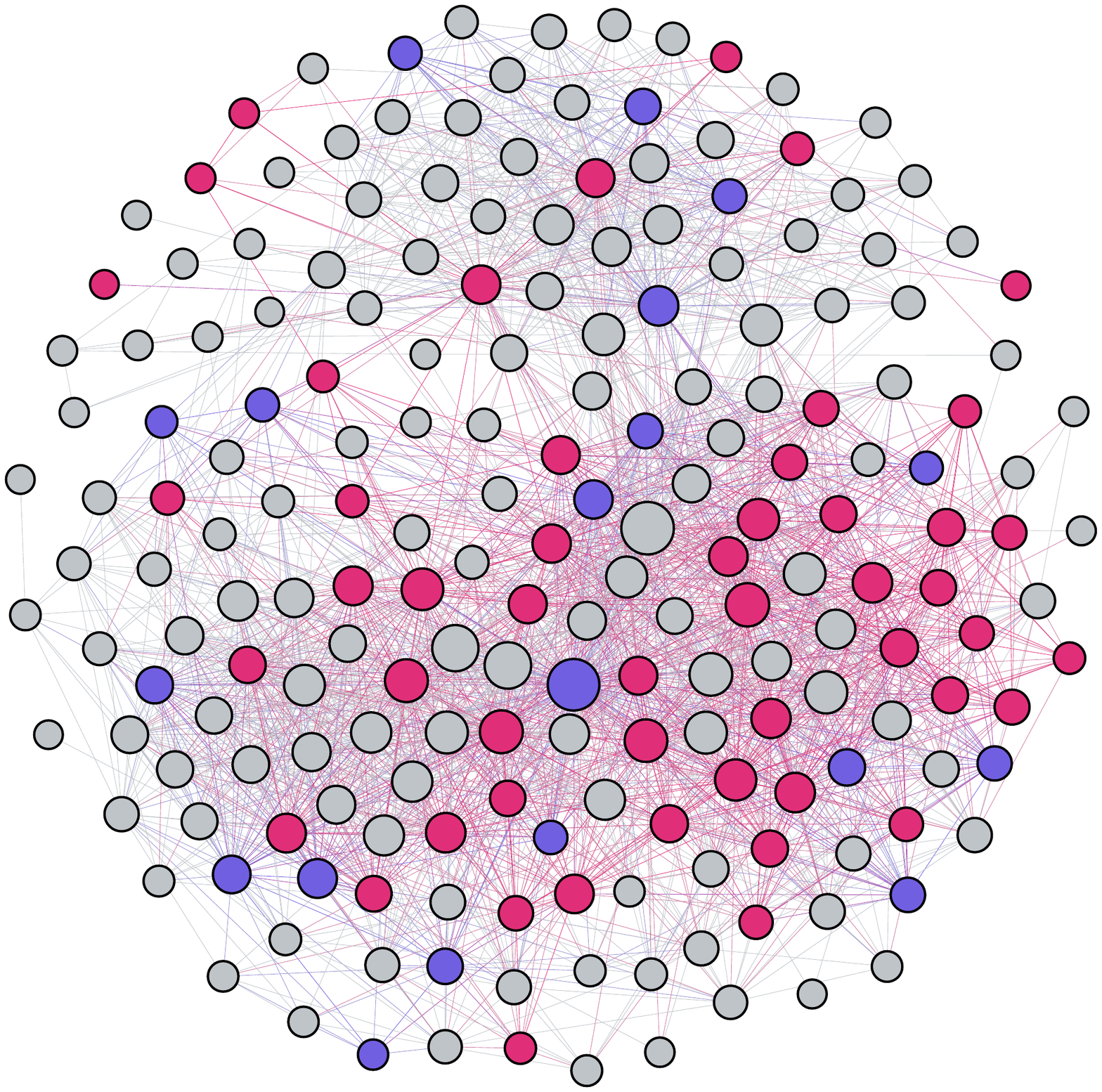}}
        \subfloat[ToupleGDD-10\%]{\label{fig: visual_sub_10}
            \hspace{-3mm}\includegraphics[width=0.2\textwidth, trim = 0cm 5cm 0cm 5cm, clip]{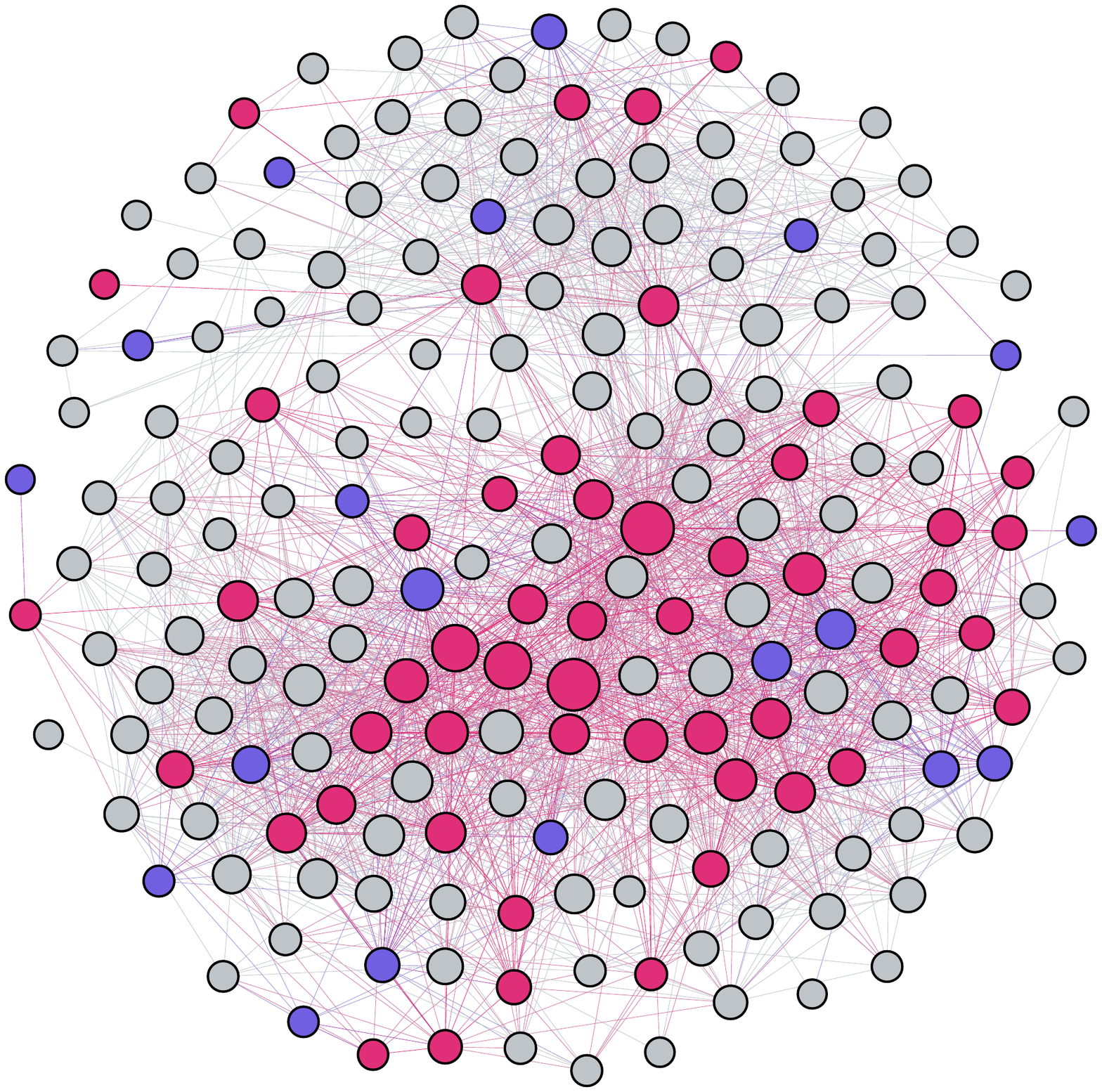}}
        \vspace{-4mm}
        \subfloat[DeepIM-20\%]{\label{fig: visual_DeepIM_20}
            \hspace{-3mm}\includegraphics[width=0.2\textwidth, trim = 0cm 5cm 0cm 4cm, clip]{figures/IC_visualization/genim_20.pdf}}
        \subfloat[OIM-20\%]{\label{fig: visual_greedy_20}
            \hspace{-3mm}\includegraphics[width=0.2\textwidth, trim = 0cm 5cm 0cm 4cm, clip]{figures/IC_visualization/oim_20.pdf}}
        \subfloat[OPIM-20\%]{\label{fig: visual_imm_20}
            \hspace{-3mm}\includegraphics[width=0.2\textwidth, trim = 0cm 5cm 0cm 4cm, clip]{figures/IC_visualization/imm_20.pdf}}
        \subfloat[SubSIM-20\%]{\label{fig: visual_opim_20}
            \hspace{-3mm}\includegraphics[width=0.2\textwidth, trim = 0cm 5cm 0cm 4cm, clip]{figures/IC_visualization/opim_20.pdf}}
        \subfloat[ToupleGDD-20\%]{\label{fig: visual_sub_20}
            \hspace{-3mm}\includegraphics[width=0.2\textwidth, trim = 0cm 5cm 0cm 4cm, clip]{figures/IC_visualization/sub_20.pdf}}
        \vspace{-3mm}
        \caption{The visualization of influence spread in Jazz dataset: The size of nodes is determined by the node degree, and the color on nodes determines the infection status: blue means the node is in seed set, red means the node is infected, and grey means the node is not infected.}
        \vspace{-5mm}
        \label{fig: visualization_1}
    \end{figure*}

\section{More Experiment}
    \subsection{Data}
    The statistics of datasets are depicted in Table \ref{tab: dataset}, and we also provide a more detailed dataset description as follows.  \textit{1) Jazz} \cite{rossi2015nr}. This dataset is a Jazz musicians collaboration network, where each node represents a musician and each edge represents two musicians who have played together in a band. \textit{2) Cora-ML} \cite{mccallum2000automating}. This network contains computer science research papers, where each node represents a paper and each edge represents one paper cites the other one. \textit{3) Power Grid} \cite{rossi2015nr}. This is a topology network of the US Western States Power Grid. An edge represents a power supply line. A node is either a generator, a transformation, or a substation. \textit{4) Network Science} \cite{rossi2015nr}. This is a coauthorship network between scientists working on network theory, where nodes represent scientists and edges represent two scientists who have collaborated. \textit{5) Digg} \cite{panagopoulos2020multi}. A directed network of social media where users follow each other and a vote to a post allows followers to see the post. \textit{6) Weibo} \cite{panagopoulos2020multi}. A directed follower network where a cascade is defined by the first tweet and the list of retweets.

	\subsection{Hyperparameter Setting.} For each baseline, we set hyperparameters according to their original papers and  fine-tune them on each dataset. For the configuration of each diffusion model, we use a weighted cascade version of the IC model, i.e., the propagation probability $p_{u,v}=1/d^{in}_v$ ($d^{in}_v$ denotes the in-degree of node $v$) for each edge $e=(u, v)$ on graph $G$; For LT model, the threshold $\theta$ is set to be uniformly sampled from $[0.3, 0.6]$ for each node $v$; the infection probability and recovery  probability are set to be $0.001$ in the SIS model. For DeepIM, the $2$-layer GAT-structured diffusion estimation model that each layer contains $4$ attention heads and the dimension of each attention channel is $64$. Both encoder and decoder are symmetric $4$-layer MLP with hidden size $512$, $1024$, $1024$, and $1024$ for each layer, respectively. We choose Adam with learning rates $0.001$ and $0.0001$ for optimizing both Eq. \eqref{eq: learning_objective_2} and Eq. \eqref{eq: infer}, respectively. 

    \subsection{IM under Non-progressive Diffusion Model}
	We demonstrate the performance of each model under the non-progressive SIS model. As shown in Table \ref{tab: evaluation_sis}, we can observe a vast number of performance reductions regarding the final influence spread compared to Table \ref{tab: evaluation_ic} and Table \ref{tab: evaluation_lt}, which is mainly due to the SIS diffusion model assumes the nodes can switch from activated to de-activated with a certain probability. Existing models all failed to capture the intrinsically more complicated diffusion dynamics. Nevertheless, DeepIM still exhibits better results and excels over others on average $10$\% on all datasets under such circumstances. To sum up, by jointly learning the seed set representation and the end-to-end diffusion estimation model, DeepIM illustrates its robustness by the capability of adapting to various underlying diffusion patterns and producing a generally competitive and stable influence spread.

    \subsection{Case Study: Graph Diffusion Visualization}
    Finally, we conduct a case study to demonstrate the distribution of selected seed nodes as well as the final infection status of all nodes in Fig. \ref{fig: visualization_1}, where blue nodes indicate the initial seed node, red nodes indicate the infected node during the influence spread, and grey nodes represent uninfected nodes. We compare the influence spread result between different initial seed set sizes, namely $10\%$ and $20\%$. Due to the ease of representation, we only visualize the result of the Jazz dataset because of its overall smaller graph size. 
	
	DeepIM demonstrates an overall better performance in terms of spreading the influence to a great extent. Notably, it can be visually seen from Fig. \ref{fig: visual_DeepIM_10} and \ref{fig: visual_DeepIM_20} that the final influence spread achieved by DeepIM with different initial seed set sizes is also very little, which means DeepIM can attain a better result with a lower cost comparing to others. The visualization demonstrates a consistent result with Table \ref{tab: evaluation_ic}.


\end{document}